\documentclass[eqsecnum,twocolumn,tightenlines,floats,floatfix,prc,nofootinbib,preprintnumbers]{revtex4-1}

\def\bea{\begin{eqnarray}}
\def\eea{\end{eqnarray}}

\def\be{\begin{equation}}
\def\ee{\end{equation}}
\def\ba{\begin{eqnarray}}
\def\ea{\end{eqnarray}}

\usepackage{graphics}
\usepackage{graphicx}
\usepackage{epsf} 
\usepackage{amsmath}
\usepackage{amssymb}
\usepackage{slashed}
\usepackage[usenames]{color}
\definecolor{MyDarkBlue}{rgb}{0,0.08,0.45}

\usepackage{bbold}
\usepackage{bm}

\newcommand{\R}[1] {\color{red}#1} 

\def\sfrac#1#2{{\textstyle \frac{#1}{#2}}}

\begin{document}

\phantom{0}
\vspace{-0.2in}
\hspace{5.5in}

\preprint{JLAB-THY-12-1489}

\title{\bf 
Spin and angular momentum in the nucleon\\ \phantom{0}}
\author{
Franz Gross$^{1,2}$, G.~Ramalho$^{3}$ 
and
M.~T.~Pe\~na$^{3}$
\vspace{-0.1in}  }

\affiliation{
$^1$Thomas Jefferson National Accelerator Facility, Newport News, 
Virginia 23606, USA \vspace{-0.15in}}
\affiliation{
$^2$College of William and Mary, Williamsburg, Virginia 23185, USA
\vspace{-0.15in}}
\affiliation{
$^3$Universidade T\'ecnica de Lisboa, CFTP, Instituto Superior T\'ecnico,
Av.\ Rovisco Pais, 1049-001 Lisboa, Portugal}

\phantom{0}

\begin{abstract}
Using the covariant spectator theory (CST), we present the results of a valence quark-diquark model calculation of the nucleon  structure function $f(x)$ measured in unpolarized deep inelastic scattering (DIS), and the structure functions $g_1(x)$ and $g_2(x)$ measured in DIS using polarized beams and targets.  Parameters of the wave functions are adjusted to fit all the data. The fit fixes both the shape of the wave functions and the relative strength of each component.  Two solutions are found that fit $f(x)$ and $g_1(x)$, but only one of these gives a good description of $g_2(x)$.  This  fit  requires  the nucleon CST wave functions contain a  large D-wave component (about 35\%) and a small P-wave component (about 0.6\%).  The significance of these results is discussed.
\end{abstract}

\vskip 1cm
\date{\today}
\maketitle

\vspace{-1.5cm}

\section {Introduction}

The first measurements of deep inelastic scattering (DIS) from  polarized protons produced a surprising result \cite{Ashman:1987hv,Ashman:1989ig} which became known as the proton spin crisis \cite{Ellis74,Jaffe:1989jz}:  it turned out that the structure function $g_1^p(x)$ (where $x$ is the Bjorken scaling variable, defined below) gave a result much smaller that expected.   At large  $Q^2$ (where $q^2=-Q^2$ is the square of the four momentum transferred by the scattered lepton),  recent measurements at a number of experimental facilities \cite{Kuhn:2008sy,Chen:2010qc} give 
\bea
\Gamma_1^p&=&\int_0^1 dx\,g^{\rm exp\pm}_{1p}(x)=0.128\pm 0.013
\eea
while theoretical calculations based on the naive assumption that the nucleon is made of quarks in a pure relative S-state give  much larger values (Jaffe and Manohar  \cite{Jaffe:1989jz} give 0.194, and our model gives 0.278, as discussed in Sec.\ \ref{sec:III} below.  Note that the experimental value of 0.128 is remarkably close to the older value of 0.126  \cite{Ashman:1987hv,Ashman:1989ig} cited by Jaffe and Manohar.)

The controversy was sharpened by  Ji \cite{Ji:1996ek,Ji:1997pf} who introduced a gauge invariant decomposition of the nucleon spin into spin and angular momentum components. This spin sum rule can be written \cite{spinchapter,FJ01}
\bea
\sfrac12=\sfrac12 \Sigma+L_q +J_G \label{eq:spinsum}
\eea
where $\Sigma$ is the contribution from the quark spins, $L_q$ the contribution from quark orbital momenta, and $J_G$ the total gluon contribution.  Some experimental estimates suggest that $\Sigma\simeq0.3$ requiring that most of the explanation for the proton spin come from the other contributions, but there is evidence that the gluon contributions are small and  it is unclear how to interpret this spin sum rule \cite{Bakker:2004ib}.

With this spin puzzle as background, we decided to see what our covariant constituent quark model, based on the covariant spectator theory (CST) \cite{Gross:1969rv,Gross:1972ye,Gross:1982nz}, would predict for the DIS structure functions.  This model was originally developed to describe  the nucleon form factors \cite{Gross:2006fg}, and has since been used to describe many other electromagnetic transitions between baryonic states, including the $N\to\Delta$ \cite{Ramalho:2008ra,Ramalho:2008dp}, the $\Delta$ form factors \cite{Ramalho:2008dc,Ramalho:2009vc,Ramalho:2010xj},  $N\to N^*(1440)$ \cite{Roper}, $N\to N^*(1535)$ \cite{Ramalho:2011ae}, and $N\to \Delta^*(1600)$ \cite{Ramalho:2010cw}.  All of these calculations  use constituent 
valence quarks with form factors of their own (initially fixed  by the fits to the nucleon form factors), and then model the baryon wave functions using a quark-diquark model with a few parameters adjusted to fit the form factors and transition amplitudes.   The diquarks have either spin-0 or or spin-1 with four-vector polarizations in the fixed-axis representation \cite{Gross:2008zza}.

One shortcoming of our model, as it has been applied so far, is that we have not yet included a dynamical calculation of the pion cloud.  Recently, constraints on the size of the pion cloud were obtained from a study of the SU(3) baryon octet magnetic moments \cite{Gross:2009jg,OctetFF}, and it is clear from this study that the pion cloud contributions to the nucleon form factors are not negligible.  
Furthermore, even without pion cloud effects it is difficult to untangle the form factors of the constituent quarks from the ``body'' form factors (which depend only on the wave functions of the nucleon).  We need a way to determine the nucleon wave functions independent of the contributions from the pion cloud and the constituent quark form factors.

The study of DIS provides an ideal answer to this problem.   In CST, the DIS  structure functions directly determine the valence part of the nucleon wave functions, giving both their shape and their orbital angular momentum content.  Adjusting model wave functions to fit the DIS data fixes all of these components.  The calculation of the DIS structure functions is also of great interest in itself.  With this model we can address the nucleon spin puzzle directly.
 
For this reasons we have decided to ``start over'' and  let  the 
valence part of nucleon wave functions be completely determined by a fit to the valence part of the DIS structure functions. Once the wave functions have been determined in this way, the low $Q^2$ pion cloud contributions can be calculated and the nucleon form factor data can be used to fix the only remaining unknown quantities: the constituent quark form factors.  This is planed for future work.

The remainder of this paper is divided into five sections and three Appendices.  In Sec.\ \ref{sec:qdis}, the DIS cross section and structure functions are defined, and theoretical results for the DIS structure functions are reported (the detailed calculations of these structure functions are given in Appendix \ref{app:sf}).  The calculations use nucleon wave functions defined in the accompaning paper \cite{previous}, and summarized in Sec.\  \ref{sec:IIB}.  This section also discusses how the wave functions are related to the structure functions.  In Sec.\ \ref{sec:III} the data is discussed and it is shown how a model without angular momentum components fails.  Then, in Sec.\ \ref{sec:IV}, we show how  either P or D-state components can fix $g_1^p(x)$, and a detailed fit to both the unpolarized structure function $f(x)$ and the polarized structure function $g_1(x)$ is given.  We find two solutions, but in Sec.\ \ref{sec:g2} we show that only one of them gives a good account of the smaller transverse polarization function $g_2(x)$.   Our conclusions are presented in Sec.\ \ref{sec:V}.  Some other details are discussed in the remaining Appendices.  
 
\begin{figure}
\centerline{
\mbox{
\includegraphics[width=2.7in]{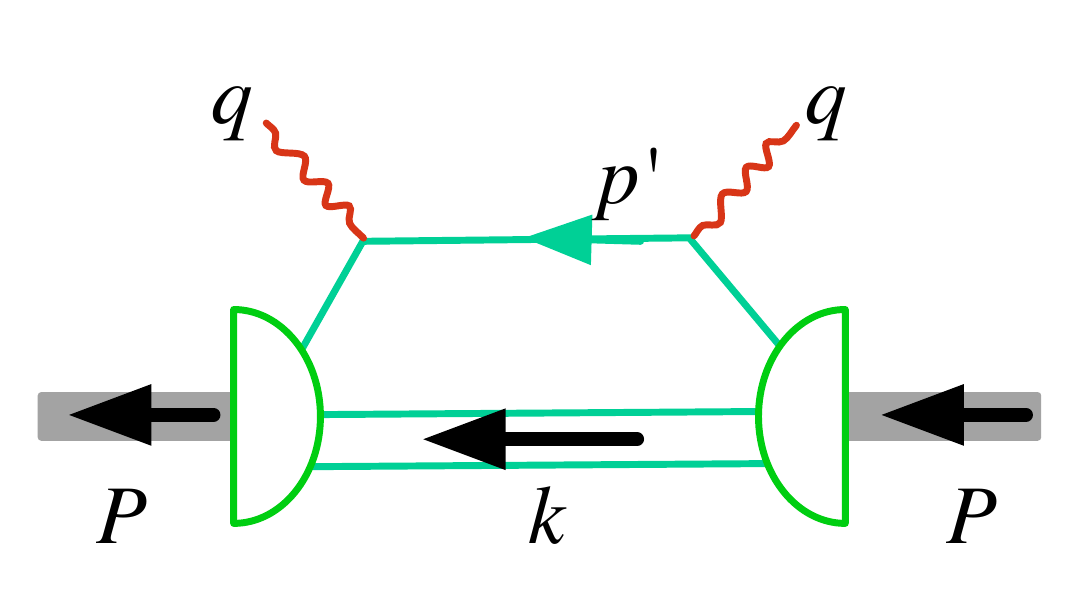} }}
\caption{\footnotesize{(Color on line) Feynman diagram for the DIS total cros section.  All of the intermediate quarks are on shell. }}
\label{fig:DIS}
\end{figure}

\section{Structure functions for DIS} \label{sec:qdis}

\subsection{Cross section in the CST}

The DIS cross section can be calculated from the imaginary part of the forward handbag diagram shown in Fig.~\ref{fig:DIS}.  The cross section depends on the hadronic tensor \cite{Kuhn:2008sy,Carlson:1998gf}
%
\bea
W_{\mu\nu}(q,P)&=&3 \sum_{s_q,\Lambda}  
\int\int \frac{d^3p'\,d^3 k}{(2\pi)^6 2E_s}\frac{m_q}{e_q}
 \nonumber\\
 &&\times 
(2\pi)^4\delta^4(p'+k-q-P)(J^{s_q}_{\Lambda\lambda})_\nu^\dagger (J^{s_q}_{\Lambda\lambda})_\mu
\cr
&\equiv&2\pi \Bigg\{\left(\frac{q_\mu q_\nu}{q^2}-g_{\mu\nu}\right)W_1+\widetilde P_\mu \widetilde P_\nu \frac{W_2}{M^2}
\nonumber \\
& & \qquad- {\cal I}_1 \Big(G_1+ G_2\frac{P\cdot q}{M^2}\Big)+{\cal I}_2 G_2
 \Bigg\} \label{hadcurr}
\eea  
%
where $P, q$ are the  four-momenta of the nucleon and the virtual photon, respectively, 
\bea
\widetilde P_\mu&&= P_\mu -
\frac{P\cdot q\,q_\mu}{q^2}
\nonumber\\
{\cal I}_1&&=\frac1{M} i \varepsilon_{\mu \nu \alpha \beta} \,q^\alpha S^\beta
\nonumber\\
{\cal I}_2&&=\frac1{M^3}(S \cdot q)\,i \varepsilon_{\mu \nu \alpha \beta} \,q^\alpha P^\beta,
\eea
$e_q=\sqrt{m_q^2+{\bf p}'^2}$ is the energy of the on-shell quark in the final state,  $E_s=\sqrt{m_s^2+ {\bf k}^2}$ is the energy of the on-shell diquark with mass $m_s$, and  $W_1, W_2, G_1$, and $G_2$ are the DIS structure functions.  
Also in Eq.~(\ref{hadcurr}) $s_q, \Lambda$ and $\lambda$ are spin projections
of the quark, the diquark and the nucleon, respectively, while $S$ stands for the nucleon four-vector polarization.

The hadronic current $(J^{s_q}_{\Lambda\lambda})_\mu$ is 
\bea
(J^{s_q}_{\Lambda\lambda})_\mu=-\bar u({\bf p}',s_q) j_\mu(q)\Psi_{\Lambda\lambda}(P,k ) 
 \label{DIScurrent}
\eea
with $u({\bf p}^\prime,s_q)$ the Dirac spinor for the quark, and 
$j_\mu(q)$ the elementary quark current with a gauge invariant subtraction
\bea
j_\mu(q)=j_q\Big(\gamma_\mu-\frac{\slashed{q}\,q_\mu}{q^2}\Big) \label{eq:qcurrent-2}
\eea
and $j_q$ the quark charge operator and $\Psi_{\Lambda \lambda}$ the nucleon wave function, with the general form
\bea
\Psi_{\Lambda \lambda}={\cal O}_\Lambda \,u(P,\lambda) \label{eq:Olambda}
\eea
to be specified below.  We emphasize that, in this context, the choice of the current (\ref{eq:qcurrent-2}) is purely phenomenological, but at least it has been shown in one special case \cite{Batiz:1998wf} that the subtraction term $\slashed{q} q_\mu/q^2$ arises naturally from interaction currents neglected here.  Once the form (\ref{eq:qcurrent-2}) is assumed, it is not necessary to explicitly calculate the contributions from the subtraction terms $\slashed{q} q_\mu/q^2$ because they can be reconstructed from the $\gamma_\mu$ term, as discussed in Appendix \ref{app:sf}.  The overall factor of 3 multiplying the hadronic tensor arises from the contributions of the three quarks \cite{Gross:2006fg,previous}.  Note that we do {\it not\/} average over the spin projections of the target nucleon; this leads to the inclusion of the last two terms in the hadronic tensor  that arise when the nucleon target is polarized.

The polarization of the nucleon is described by the four-vector polarization $S$ with the properties
\bea
S_\mu S^\mu=-1,\quad S\cdot P=0\, .
\eea
For a nucleon at rest, this polarization vector is
\bea
S=\left[\begin{array}{c}
 0 \cr
 \sin\theta\cos\phi\cr
 \sin\theta\sin\phi\cr
 \cos\theta
 \end{array}\right]\, .
\eea
We use the helicity basis to describe the nucleon spin; for a nucleon at rest we will choose the spin axis to be in the  
$+\hat z$ direction,  
(the direction of the three-vector ${\bf q}$).  
A nucleon polarized in the direction of $S$ can be written as a linear combination of helicity states, so that 
\bea
u(P,{\bf S}_m)={\cal D}_{\lambda, m}^{(\frac12)}(\phi,\theta,0)\,u(P,\lambda)
\eea
where ${\cal D}$ is the rotation matrix
\bea
{\cal D}^{(\frac12)}(\phi,\theta,0)=\left[\begin{array}{rr}
 \cos\sfrac12\theta\, e^{-i\frac\phi2} & \quad-\sin\frac12\theta \, e^{-i\frac\phi2}  \\[0.1in]
\sin\frac12\theta \, e^{i\frac\phi2}  &
 \cos\sfrac12\theta  \,e^{i\frac\phi2}  \cr
 \end{array}\right]\quad
\eea
and sum over $\lambda$ is implied.  Note that
\bea
({\bf \sigma}\cdot{\bf S})\, u(P,{\bf S}_m)=2m\, u(P,{\bf S}_m)\, ,
\eea
where $m=\pm\frac12$ is the spin projection.

Using the identity
\bea
u(P,{\bf S}_m)\bar u(P,{\bf S}_m)=\Lambda_M(P)\,\sfrac12\Big[1+\gamma^5\slashed{S}\Big]
\eea
where
\bea
\Lambda_M(P)=\frac{M+\slashed{P}}{2M}=\sum_{s}u({\bf P},s)\bar u({\bf P},s)
\eea
is the positive energy projection operator,
and summing over the spins of the outgoing quark allows the hadronic tensor to be expressed as a trace
%
\bea
&&W_{\mu\nu}(q,P)=3\sum_{\Lambda}\int\int \frac{d^3p'\,d^3 k}{(2\pi)^2 2E_s}\frac{m_q}{e_q}
\,\delta^4(p'+k-q-P)
\nonumber\\
&&\quad\times{\rm tr} \Big[{\cal O}_\Lambda j_\nu(q)\Lambda_{m_q}(p')j_\mu(q){\cal O}_\Lambda\Lambda_M(P)\sfrac12[1+\gamma^5\slashed{S}]\Big]
 \label{hadcurr-1}
\eea
%
where the operators ${\cal O}_\Lambda$ 
are defined by Eq.\ (\ref{eq:Olambda})  with the wave function spin components given explicitly in the next subsection.

\subsection{Wave function of the nucleon} \label{sec:IIB}

In this section we summarize our model of the nucleon wave function, which is composed of S, P, and D-state components.  For details see Ref.~\cite{previous}, hereinafter referred to as Ref.~I.  
Briefly, the wave function has a quark-diquark structure, with the $i$th quark off-shell (where $i=\{1,2,3\}$) and the other two {\it noninteracting on-shell\/}  quarks treated as a diquark system with total four momentum $k_i$ and fixed (average) mass $m_s$.  The total momentum of the nucleon is $P$.

The CST wave function of the nucleon is the superposition of a leading S-state component, with smaller  P and D-state components
\bea
\Psi_{\Lambda\lambda}(P,k ) &=& 
n_S\Psi^{S}_{\Lambda\lambda}(P,k)
\nonumber\\
&&\quad+
n_P \Psi^{P}_{\Lambda\lambda}(P,k)+n_D\Psi^{D}_{\Lambda\lambda}(P,k )\, . \qquad\quad\label{psiRel} 
\eea
Each component of the wave function is normalized to the same value [see Eq.\ (\ref{eq:normall}) below], so if the coefficient $n_S$ of the S-state is fixed by the coefficients $n_P$ and $n_D$  
\bea
n_S=\sqrt{1-n_P^2-n_D^2}\, ,
\eea
Then the square of each coefficient, $n_L^2$ (where $L=\{S,P,D\}$), is proportional to the percentage of each component.  The size of the coefficients  $n_P$ and $n_D$ will be fixed by the fits.  
The construction of these wave functions is discussed in detail in Ref.~I. 

The formulae are first derived under the assumption that isospin is an exact symmetry.  The formulae are then generalized  to allow for the $u$ and $d$ quark distributions to differ, and all fits were done adjusting the $u$ and $d$ quark distributions independently. The S and P-state components are a sum of terms with spin 0 (and isospin 0)  and spin 1 (and isospin 1) diquark contributions, while only diquarks of spin 1 can contribute to the D-state component.  The diquarks of spin 0 do not interfere with diquarks of spin 1.  The individual components are denoted $\Psi^{L,n}$ with $L=\{S, P, D\}$ the angular momentum and $n=\{0,1,2\}$ labeling the state of the diquark (sometimes the spin=isospin, or as in the case of the D-state, all three states have diquarks with spin 1).  These are
\bea
\Psi_{\lambda}^{S,0}&&=\sfrac1{\sqrt{2}}\,{\bm\phi}^0\;u({\bf P},\lambda) \psi_S(P,k)
\nonumber\\
\Psi_{\Lambda\lambda}^{S,1}&&=-\sfrac1{\sqrt{2}}\,{\bm\phi}^1\,(\varepsilon^*_\Lambda)^\alpha U_\alpha({\bf P},\lambda) \psi_S(P,k)
\nonumber\\
\Psi_{\lambda}^{P,0}&&=\sfrac1{\sqrt{2}}\,{\bm\phi}^0\;\slashed{\widetilde k}\;u({\bf P},\lambda) \psi_P(P,k)
\nonumber\\
\Psi_{\Lambda\lambda}^{P,1}&&=-\sfrac1{\sqrt{2}}\,{\bm\phi}^1\;\slashed{\widetilde k}\;(\varepsilon^*_\Lambda)^\alpha U_\alpha({\bf P},\lambda) \psi_P(P,k)
\nonumber\\
 \Psi^{D,0}_{\Lambda \lambda}   &&=
\sfrac{3}{2\sqrt{10}}\,{\bm\phi}^0\, ({\varepsilon}^*_\Lambda)_\alpha
G^{\alpha\beta}(\tilde k, \zeta_\nu)
U_\beta({\bf P},\lambda)|\tilde k|\psi_D(P,k) 
\qquad 
\nonumber\\
 \Psi^{D,1}_{\Lambda\lambda}  &&=-\sfrac{1}{\sqrt{10}}\,{\bm\phi}^1 \,
{\epsilon}_{D\Lambda}^{\beta *}U_\beta({\bf P},\lambda)\,\tilde k^2\psi_D(P,k) 
\nonumber\\
\Psi_{\Lambda\lambda}^{D,2}&&=\sfrac3{\sqrt{5}}\,{\bm\phi}^1\,(\varepsilon^*_\Lambda)_\beta\, D^{\beta\alpha}(P,k) U_\alpha({\bf P},\lambda) \psi_D(P,k)\qquad\quad
\eea
where ${\bm\phi}^I$ (with $I=0$ or 1) are the isospin $I$ parts of the wave function (discussed in the next subsection), $\psi_L(P,k)$ are the scalar $S, P,$ or $D$ wave functions, $\varepsilon^*_\Lambda$ is the {\it outgoing\/} spin one diquark  four-vector polarization with spin projection $\Lambda$ in the direction of ${\bf P}$, ${\epsilon}_{D\Lambda}^{*}$ is the outgoing four-vector diquark with an internal D-wave structure and spin projection $\Lambda$ in the direction of ${\bf P}$, and the other four-momenta and operators are
\bea
&&\widetilde k=k-\frac{(P\cdot k)P}{M^2}
\nonumber\\
&&\widetilde\gamma_\alpha=\gamma_\alpha-\frac{\slashed{P}P_\alpha}{M^2}
\nonumber\\
&&U_\alpha({\bf P},\lambda)=\sfrac1{\sqrt{3}}\gamma^5\widetilde \gamma_\alpha u({\bf P},\lambda)
\nonumber\\
&&G^{\alpha\beta}(\widetilde k, \zeta_\nu)=\widetilde k^\alpha \zeta_\nu^\beta+ \zeta_\nu^\alpha \widetilde k^\beta-\sfrac23 \tilde g^{\alpha\beta} (\widetilde k\cdot \zeta_\nu)
\nonumber\\[0.05in]
&&D_{\beta\alpha}(P,k)=\widetilde k_\beta \widetilde k_\alpha-\sfrac13 \widetilde g_{\beta\alpha}\,\widetilde k^2
\nonumber\\
&&\widetilde g_{\beta\alpha}=g_{\beta\alpha}-\frac{P_\beta P_\alpha}{M^2}\, , \label{eq:2.15}
\eea
with $G(\widetilde k, \zeta_\nu)$ the tensor describing the spin-two coupling of a diquark with an internal P-wave orbital angular momentum structure to the P-wave motion of the third, off-shell quark, and $D(P,k)$ the spin-two tensor describing the a D-wave motion of the off-shell quark.  Note that $P$ is orthogonal to all of the polarization vectors, and $P\cdot \widetilde k= P\cdot \widetilde \gamma=0$, implying that 
$P^\alpha U_\alpha(P,\lambda)=0$, and  
\bea
P^\beta D_{\beta\alpha}&&=P^\alpha D_{\beta\alpha}=0
\nonumber\\
P_\beta G^{\beta\alpha}&&=P_\alpha G^{\beta\alpha}=0\, . 
\label{eq:pdotG}
\eea

As discussed in Ref.~I, when isospin is conserved, the S, P and D-state wave functions are normalized to
\bea
1&&=e_0\int_k |\psi_S(P,k)|^2=e_0\int_k (-\tilde k^2) |\psi_P(P,k)|^2
\nonumber\\
&&=e_0\int_k \tilde k^4|\psi_D(P,k)|^2 \label{eq:normall}
\eea
where $e_0$ is the charge of the dressed quark at $Q^2=0$.  With this normalization the square of the coefficients $n_D$ and $n_P$ can be interpreted as the fraction of the total wave functions consisting  of D and P-state components.

\subsection{Quark charge operators in flavor space}

\subsubsection{Isospin symmetry}
 
If isospin is a good summetry, the quark charge operator in flavor space is
\bea
j_q=\sfrac16+\sfrac12\tau_3.
\eea
The isospin operators ${\bm\phi}^I$ are
\bea
{\bm\phi}^0&&={\bf 1}
\nonumber\\
{\bm\phi}^1_\ell&&=-\sfrac1{\sqrt{3}}\,\tau\cdot \xi^*_\ell 
\eea
where $\xi_\ell$ is the isospin three-vector of the diquark with isospin projection $\ell$.  These are operators; to convert them to isospin states of the off-shell quark they are multiplied from the right by the $\chi_t$, the isospin 1/2 spinor of the nucleon (see Ref.~I).   
The isospin operators are normalized to
\bea
({\bm \phi}^0)^\dagger{\bm \phi}^0=&&{\bf 1}
\nonumber\\
\sum_\ell({\bm \phi}^1_\ell)^\dagger {\bm \phi}^1_\ell=&&\sfrac13\sum_\ell\tau\cdot\xi_\ell\;\tau\cdot\xi^*_\ell={\bf 1}
\eea

The matrix elements of the quark charge operators in DIS are now easily evaluated.  Noting that DIS involves the {\it square\/} of the quark charge, the result is (including the overall factor of 3)
\bea
3\,({\bm \phi}_0)^\dagger\, j_q\,j_q \,{\bm \phi}_0&=& \sfrac56+\sfrac12\tau_3\equiv(e_q^2)^0\nonumber\\
3\sum_\ell ({\bm \phi}_\ell^1)^\dagger\, j_q\,j_q \,{\bm \phi}_\ell^1&=&\sfrac56-\sfrac16\tau_3\equiv(e_q^2)^1. \label{eq2.11b}
\eea
%


\subsubsection{Broken isospin}
 
So far this discussion assumes that the $u$ and $d$ distributions are identical, with their relative contributions being fixed only by isospin invariance.  In fact these distributions are quite different at both low and high $x$, and we know that the angular momentum distributions of the $u$ and $d$ quarks are also quite different. 

Using the results of Ref.~I for broken isospin, Eq.~(6.4),  new effective charge operators (including the overall factor of 3) are introduced
\bea
(e_q^2)^0&=&\sfrac56+\sfrac12\tau_3
\nonumber\\
(e_q^2)^{1}_0&\to&\sfrac13(\tau\cdot\xi_0)(\sfrac56+\sfrac12 \tau_3) (\tau\cdot\xi^*_0)=\sfrac13(\sfrac56+\sfrac12 \tau_3) 
\nonumber\\
(e_q^2)^{1}_1&=&\sfrac13(\tau\cdot\xi_+)(\sfrac56+\sfrac12 \tau_3) (\tau\cdot\xi^*_+)
\nonumber\\&&
+\sfrac13(\tau\cdot\xi_-)(\sfrac56+\sfrac12 \tau_3) (\tau\cdot\xi^*_-)
\nonumber\\
&=&\sfrac23(\sfrac56-\sfrac12 \tau_3)
\eea
where it is understood that $(e_q^2)^0$ and $(e_q^2)^{1}_0$ multiply $u$ quark distributions and $(e_q^2)^{1}_1$ multiplies $d$ quark distributions.  In order to simplify the notation, we write these for the {\it proton only\/}; the neutron is obtained from the proton by charge symmetry, substituting $e_u\leftrightarrow e_d$.
Hence
\bea
(e_q^2)^0&=&3 e_u^2
\nonumber\\
(e_q^2)^{1}_0&=&  e_u^2 
\nonumber\\
(e_q^2)^{1}_1&=& 2 e_d^2
\eea
where $e_u=2/3$, $e_d=-1/3$.  Hence
\bea
(e_q^2)^0 \psi_{L'} \psi_L&=& 3\, e_u^2 \psi_u^{L'}\psi_u^L
\nonumber\\
(e_q^2)^1 \psi_{L'} \psi_L&\to&  e_u^2 \psi_u^{L'}\psi_u^L + 2  e_d^2 \psi_d^{L'}\psi_d^L
 \label{eq:2.24}
\eea
with $\psi_u^L$ and $\psi_d^L$ are the $u$ and $d$ distributions for the $L=\{S,P,D\}$ state components (see Ref.~I).  

We are now ready to report the final results for the structure functions.


\subsection{Results for the structure functions}\label{sec:IID}

Details of the calculations of the structure functions are reported in Appendix \ref{app:sf}.  
All of the results can be expressed in terms of the following functions, which depend on the quark flavor $q=\{u,d\}$ and, in some cases, on $L$, which is either the letter $\{S, P, D\}$  or the number $L=\{0,1,2\}$ :
 \bea
 f_q^L(x)&\equiv&\int_{\chi} k^{2L}[\psi_q^L(\chi)]^2
 \nonumber\\
g^{L}_q(x)&=&\int_{\chi}P_2(z_0)\,k^{2L}[\psi^L_q(\chi)]^2 \qquad  L\ge 1 \;{\rm only}
\nonumber\\
 d_q(x)&\equiv& \int_{\chi}P_2(z_0)k^2\,\psi_q^S(\chi)\psi_q^D(\chi)\, , \label{kpintb}
 \eea
 and for $L=\{0,2\}$ only, 
 \bea
 h^{L}_q(x)&\equiv& \int_{\chi}  k^{L+1}\,z_0\,\psi_q^{L}(\chi)\psi_q^P(\chi)  
 \nonumber\\
 h^{L+1}_q(x)&\equiv& \int_{\chi} \frac{k^{ L+2}}{4Mx}(1-z_0^2)\,\psi_q^L(\chi)\psi_q^P(\chi)\, , \qquad 
\label{kpinta}  
 \eea
%
%
where the integral is
\bea
\int_{\chi}\equiv  \frac{Mm_s}{16\pi^2 }\int_\zeta^\infty d\chi , \label{eq:CSTint}
\eea
and in these expressions, $k=|{\bf k}|$ is the magnitude of the three-momentum of the spectator diquark (distinguished from the four-momentum $k$ only by context), and $z_0$ is the cosine of the scattering angle fixed by the DIS condition; see Eq.\ (\ref{eq:z0}) below.  
The function $P_2(z)$ is the Legendre 
polynomial $P_2(z)=\frac{1}{2}(3 z^2-1)$.
Since the wave functions depend on only one function  $\chi$ of the four momenta $P$ and $k$ [defined and discussed below in Eqs.\ (\ref{eq:chi}) and (\ref{eq:chiCST})], we use the notation $\psi(P,k)\equiv\psi(\chi)$ for convenience.   
The physical interpretation of these expressions will  be discussed in  Sec.\ \ref{sec:IIIA}.

In terms of these structure functions, the results for the DIS observables  for the proton (with the neutron results obtained by the substitution $e_u\leftrightarrow e_d$) are:
\bea
\nu W^p_2 =&& 2MxW^p_1
=x \,2e_u^2\,f_u
+x\,e_d^2\,f_d
=xf_p(x)
\nonumber\\
g^p_1(x)=&&
e_u^2\,\Big[\sfrac23(f_u-\sfrac43\,n_P^2\,f_u^P)-n_D^2\,f_u^D\Big]
\nonumber\\
&&
-e_d^2\,\sfrac16\Big[f_d-\sfrac43\,n_P^2\, f_d^P\Big] +n_P^2\sfrac23\,G_P^p+n_D^2\sfrac23\,G_D^p
\nonumber\\
&& -\sfrac29\,a_{SD}[e_u^2\,d_u+2e_d^2\,d_d] 
\nonumber\\[0.05in]
&&+
\sfrac29\,a_{PD}[e_u^2 h_u^2+2\,e_d^2\,h_d^2]
\nonumber\\[0.05in]
g^p_2(x)=&&-n_P^2\,G_P^p-n_D^2\,G_D^p
+\sfrac13\,a_{SD} [e_u^2\,d_u+2e_d^2\,d_d]
\nonumber\\
&&
-\sfrac13 n_Pn_S\Big[4\,e_u^2[h_u^1-h_u^0] 
-e_d^2[h_d^1-h_d^0]\Big]
\nonumber\\
&&+\sfrac29\,a_{PD}\Big[e_u^2[h_u^3-h_u^2] 
+2\,e_d^2[h_d^3-h_d^2]\Big]
\label{eq:335aa}
\eea
where
\ba
& &
g_1= \frac{M}{P \cdot q} G_1 \nonumber \\
& &
g_2= \frac{M^3}{(P \cdot q)^2} G_2\, ,
\nonumber 
\ea
we used the shorthand notation $z=z(x)$ (where $z$ is any of the structure functions),  
\bea
f_q&=&n_S^2f_q^S-2n_Pn_S\,h^0_q+n_P^2\,f_q^P+n_D^2\,f_q^D
\nonumber\\
G^p_P&=&\sfrac1{3}[4\,e_u^2\,g^P_u-e_d^2\,g^P_d]
\nonumber\\
G^p_D&=&\sfrac1{40}[29\,e_u^2\,g^D_u+16\,e_d^2\,g^D_d]\, , \label{eq:fq}
\eea
and, for convenience, we introduce the coefficients
\bea
a_{SD}&=&-3\sqrt{\sfrac25}\,n_S\,n_D
\nonumber\\
a_{PD}&=&-3\sqrt{\sfrac25}\,n_P\,n_D\label{eq:ad}
\eea
to describe the strength of the SD and PD interference terms.  These results are a summary of the detailed calculations leading to Eqs.~(\ref{eq:B23}), (\ref{eq:335}), (\ref{eq:B42}), (\ref{eq:sdinter}), and (\ref{eq:B42a}). 

The formulae (\ref{eq:335aa}) can be separated into separate $u$ and $d$ contributions using the general relations
\bea
f_n(x)&=&\sum_q e_q^2\, f_q(x)
\nonumber\\
g^n_i(x)&=&\sfrac12\sum_q e_q^2\, g_i^q(x)\, \label{eq:polex}
\eea
with $i=\{1,2\}$.   Limiting the expansions to $u$ and $d$ quarks,  and ignoring antiquark, gluon, and correction terms coming from the QCD evolution, the extracted $f_q$ distributions were given in Eq.\ (\ref {eq:fq}).  The results for the $g$'s are 
\bea
 g_1^u&=&
 \sfrac23 f_u-n_D^2\,f_u^D-\sfrac89\,n_P^2\,f_u^P
 \nonumber\\
 &&  -\sfrac29\,a_{SD}\,d_u +\sfrac29\,a_{PD}\,h^2_u 
+\sfrac89\, n_P^2\,g_u^P+\sfrac{29}{60} n_D^2\,g_u^D
\nonumber\\
g_1^d&=&
-\sfrac13 f_d+\sfrac49\,n_P^2\,f_d^P 
\nonumber\\
&&-\sfrac89 a_{SD}\,d_d  +\sfrac89 a_{PD}\,h^2_d
-\sfrac49\,n_P^2\,g_d^P+ \sfrac8{15} n_D^2\,g_d^D
\nonumber\\
g_2^u&=&\sfrac13 a_{SD}d_u-\sfrac43\,n_Pn_S (h_u^1-h_u^0)+\sfrac29 a_{PD}(h_u^3-h_u^2) 
\nonumber\\
&&-\sfrac43\, n_P^2\,g_u^P-\sfrac{29}{40} n_D^2\,g_u^D
\nonumber\\
g_2^d&=&\sfrac43\,a_{SD} d_d +\sfrac23 n_Pn_S (h_d^1-h_d^0)+\sfrac89 a_{PD}(h_d^3-h_d^2)
\nonumber\\
&&+\sfrac23\,n_P^2\,g_d^P- \sfrac45 n_D^2\,g_d^D\, .\qquad  \label{eq:theoryquark}
\eea

We conclude this section with a discussion of the interpretation and normalization of the structure functions.

 \subsection{Physical interpretation}\label{sec:IIIA}


As discussed in Ref.\ I \cite{previous}, the wave functions are chosen to be simple functions of the covariant variable
\bea
\chi=\frac{(M-m_s)^2-(P-k)^2}{Mm_s}=\frac{2P\cdot k}{Mm_s}-2\, .\qquad \label{eq:chi}
\eea
Since the nucleon and the diquark are both on shell, the variable $P\cdot k$, related to the square of the mass $(P-k)^2$ of the off-shell quark, is the only possible variable on which the scalar parts of the wave functions can depend, and $\chi$ is simply a convenient linear function of this  variable.

For DIS studies we choose to work in the rest frame of the nucleon (but our results are frame independent).  In variables natural to the CST, $\chi$ depends only on the magnitude of the spectator three-momentum, which will be scaled by the mass of the nucleon.  If 
$r$ is the mass ratio
\be
r=\frac{m_s}{M}, \label{ratio}
\ee
then  $k\equiv M \kappa$, $E_s=ME_\kappa=M\sqrt{r^2+\kappa^2}$, and
\bea
\chi= \chi(\kappa)&&=\frac{2ME_\kappa}{m_s}-2=2\sqrt{1+\frac{\kappa^2}{r^2}}-2\, . \qquad\label{eq:chiCST}
\eea

\begin{figure}
\leftline{
\mbox{
\includegraphics[width=2.8in]{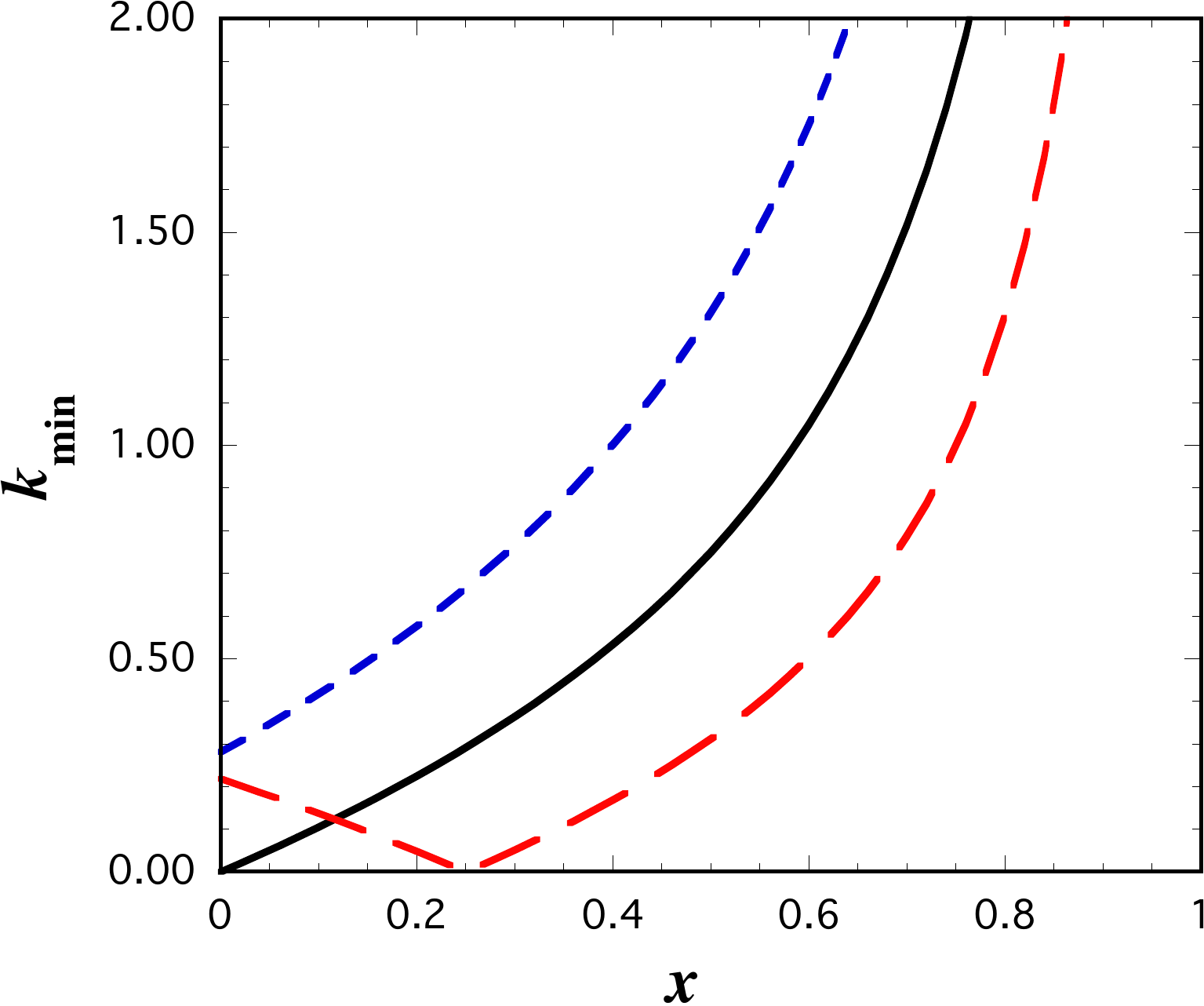}}}
\caption{\footnotesize{(Color on line) The boundary of the $\kappa$ integral, $\kappa_{\rm min}$, as a function of $x$ for selected values of $r=1.25$ (short-dashed line), $r=1$ (solid line), $r=0.75$ (long dashed line). }}
\label{fig:boundary}
\end{figure}

The structure functions, displayed in Eqs.\ (\ref{kpintb}) --
(\ref{eq:CSTint}) (discussed in Appendix \ref{app:sf}), are integrals over the magnitude of the scaled three-momentum $\kappa$, or alternatively integrals over $\chi$.  In the nucleon rest frame this integral has the form
\bea
&&\int \frac{Md^3 k}{(2\pi)^32E_s}\delta\left(E_s-k \cos\theta-M(1-x)\right) \psi^2(\chi)\qquad
\nonumber\\
&&\qquad=\frac{M^2}{(2\pi)^2}\int_0^\infty\frac{ \kappa d\kappa}{2E_\kappa}\int_{-1}^1\kappa\,dz\,\delta(E_\kappa-\kappa z-1+x)\psi^2(\chi)
\nonumber\\
&&\qquad=\frac{M^2}{(2\pi)^2}\int_{\kappa_{\rm min}}^\infty\frac{ \kappa d\kappa}{2E_\kappa}\psi^2(\chi)
\eea   
where the $\delta$ function in the last integral fixes  the scattering angle $z=\cos\theta$ in terms of the momentum $\kappa$ and the Bjorken variable $x$  [the CST form of the DIS scattering condition, see Eq.\ (\ref{DISint-a})] 
\bea
z\to z_0&=&\frac{E_\kappa-(1-x)}{\kappa}
\nonumber\\
&=&\frac{r\chi +2(r-1+x)}{r\sqrt{\chi(\chi+4)}}
 \, .\label{eq:z0} 
\eea
The requirement that the scattering angle be physical, or that $|z_0|\leq1$, fixes the lower limit of the $\kappa$ integration at
\bea
\kappa&\ge&|\kappa_{\rm min}|
\nonumber\\
\kappa_{\rm min}&\equiv&\frac{r^2-(1-x)^2}{2(1-x)}
\eea
for $x\in [0,1]$.    The boundary of the region of integration over $\kappa$, $|\kappa_{\rm min}|$, is shown in Fig. \ref{fig:boundary} for selected values of the ratio $r$.  When $r<1$, the boundary has a cusp at $x=1-r$, which deserves further comment.  At the cusp $|\kappa_{\rm min}|=0$; if $x>1-r$ the region $\kappa<|\kappa_{\rm min}|$ is excluded because $z_0>1$ while if $x<1-r$ the region $\kappa<|\kappa_{\rm min}|$ is excluded because $z_0<-1$.  In either case the lower limit of $\chi$ becomes
\bea
\nonumber\\
\zeta&&=
\frac{r}{1-x}+\frac{1-x}{r}-2
=\frac{(r+x-1)^2}{r(1-x)}
\, ,\qquad \label{eq:3.13}
\eea
with (for future reference)
\bea
\frac{d\zeta}{dx}=\frac{2\kappa_{\rm min}}{r(1-x)}\, .
\eea
Hence the integral over $\kappa$ can be transformed to
\bea
\frac{M^2}{(2\pi)^2}\int_{\kappa_{\rm min}}^\infty\frac{ \kappa d\kappa}{2E_\kappa}\psi^2(\chi)=\frac{Mm_s}{4(2\pi)^2}\int_{\zeta}^\infty d\chi\,\psi^2(\chi)\qquad
\eea
leading to integrals of the form (\ref{eq:CSTint}).

It is now easy to interpret the physical meaning of the structure functions in the context of the CST.  The formula 
\bea
f_q^S(x)&&=\frac{Mm_s}{16\pi^2}\int_{\zeta}^\infty d\chi \,[\psi_q^S(\chi)]^2 \label{eq:3.7}
\eea
displays the simplest structure function as an average over the square of the wave function $[\psi_q^S]^2$ with a variable lower limit that depends on $x$.   The wave function can be unfolded from this average by differentation
\bea
[\psi_q^S(\zeta)]^2&=&-\frac{r(1-x)}{2\kappa_{\rm min}}\frac{16\pi^2}{m_sM}\frac{df^S_q(x)}{dx} 
\nonumber\\
&=& -\frac{(1-x)^2}{x(2-x)}\frac{16\pi^2}{m_sM}\frac{df^S_q(x)}{dx} \quad{\rm if}\;r=1\, .\qquad\quad
\label{eq:3.21}
\eea

Examination of the experimentally determined distributions, discussed in Sec.\ \ref{sec:III} below, {\it shows that the derivative is singular at $x=0$, and is nonzero throughout the region $0\le x\le 1$\/}.  Therefore, the wave function must have a singularity at $x=0$ corresponding to $\kappa=\frac12|r^2-1|$.  Unless $r=1$ this singularity is is at a finite and non-vanishing value of $\kappa$, which is unphysical.   Furthermore, if we  choose $r<1$, the wave function will have another singularity at $x=1-r$, corresponding to $\zeta=0$ and $\kappa=0$.  [In Ref.\ \cite{Gross:2006fg} we used $r<1$ and a wave function that was finite at all $\kappa$, inevitably giving a distribution amplitude with the wrong shape (see Fig.\ [9] in Ref.\ \cite{Gross:2006fg}).]    We conclude that no choice of $r$ will allow us to chose a wave function that is not singular at some momentum, but choosing $r=1$ where the only singularity is at  $\kappa=0$ where Dirac wave functions are known to be singular, makes sense physically.
 With this choice the integral (\ref{eq:CSTint}) samples the wave functions over the {\it entire range of momentum $\kappa$\/}, with the sample size depending on the value of $x$, as shown in Fig.\ \ref{fig:boundary}.

Assuming for the moment that $f^S_q(x)\to x^\alpha (1-x)^\gamma$ at large and small $x$, this shows that the {\it square\/} of the S-state wave function, when $r=1$, must go as $x^{\alpha-2}(1-x)^{\gamma+1}$ at large and small $x$.  Since
\bea
\zeta=\frac{x^2}{1-x},
\eea
this behavior requires the {\it square\/} of the wave function to go, at large and small $\zeta$, like 
\bea
|\psi_q^S(\zeta)|^2\sim \frac1{\zeta^{1-\alpha/2}(\beta+\zeta)^{\gamma+\alpha/2}}\, ,\label{eq:asywf}
\eea
where $\beta$ is a range parameter. In this way the asymptotic behaviors of the wave function can be estimated directly from the structure function, at least for cases where the integrand does not depend on the cosine of the scattering angle, $z_0$.  (Recall that the argument of the wave function itself does not depend on $z_0$.)  We will use this insight in Sec.\ \ref{sec:IV} to guide our construction of models for the wave functions.

We conclude this subsection by noting that the expression (\ref{eq:3.7})  display a certain symmetry in $f_q^S$ which is most easily discussed if we introduce $y\equiv1-x$ and define
\bea
\widetilde f_q^S(y)=f_q^S(1-x)\, . \label{eq:xtoy}
\eea
Then, from the form of $\zeta$ (written here for $r=1$, but easily generalized to $r\ne1$),
\bea
\widetilde f_q^S(y)=\widetilde f_q^S\left(\frac1y\right)\,�.\label{eq:ysymm}
\eea
This symmetry allows us to extend the definition of $\widetilde f(y)$ from the interval $0\le y\le 1$ to the interval $0\le y\le \infty$, and will be used in the next section.

\subsection{Normalization}

The representation (\ref{eq:3.7}) also allows a simple expression for the first moment of $f_q^S(x)$, or $\widetilde f_q^S(y)$
\bea
\int_0^1 dx\,f^S_q(x)&&=\frac{Mm_s}{16\pi^2}\int_0^1 dx\,\int_{\zeta}^\infty d\chi \,[\psi_q^S(\chi)]^2
\nonumber\\
&&=\frac{Mm_s}{16\pi^2}\int_0^1x dx\, \frac{d\zeta}{dx} \,[\psi_q^S(\zeta)]^2
\nonumber\\
&&=\int_0^1 dy\,\widetilde f_q^S(y) \qquad\label{eq:normf}
\eea

This can be compared with the CST normalization integral for $\psi_q^S$.  Recalling that the quark charge, at $Q^2=0$ is dressed to $e^0_q$ (in units of the charge at high $Q^2$), the CST normalization integral is
\bea
1&&=e^0_q\int\frac{d^3 k}{(2\pi)^32E_s}[\psi_q^S(\chi)]^2
\nonumber\\
&&=\frac{e^0_q Mm_s}{8\pi^2}\int_{0}^\infty \kappa\, d\chi\,[\psi_q^S(\chi)]^2
 .\label{eq:normCST}
\eea
Note that both (\ref{eq:normf}) and (\ref{eq:normCST}) are integrals over the wave function, with $\chi$ a function of $\kappa$ and $\zeta$ a function of $x$.  We can transform these two integrals into the same form if we set $\zeta=\chi$, which defines  a mapping between $x$ and $\kappa$
\bea
\kappa^2=\frac{r^2}{4}\left(\frac{r}{1-x}-\frac{1-x}{r}\right)^2. \label{eq:kvsx}
\eea
Setting $r=1$ from here on, the normalization integral (\ref{eq:normCST}) is transformed into
\bea
1&=&\frac{e^0_q\,Mm_s}{16\pi^2}\int_{0}^1\frac{x(2-x)}{1-x} \,dx\,\frac{d\zeta}{dx}\,[\psi_q^S(\zeta)]^2
\nonumber\\
&=&\frac{e^0_q\,Mm_s}{16\pi^2}\int_{0}^1\Big[x+\frac{x}{1-x}\Big] \,dx\,\frac{d\zeta}{dx}\,[\psi_q^S(\zeta)]^2.\qquad
\label{eq:norm2}
\eea
The normalization integral, apart from the  quark charge renormalization $e_q^0$,  differs from the first moment of $f^S_q(x)$ in the weight function, which now includes the additional factor of $x/(1-x)$.

As an alternative to (\ref{eq:norm2}), write the integral in terms of $y$, use the symmetry property of $\zeta$ to transform the second term into an integral over $z=1/y$, and then replace the integration variable $z$ by $y$, allowing the normalization integral to be written
\bea
1&&=\frac{e_q^0\,Mm_s}{16\pi^2}\int^{\infty}_0(1-y)dy\,\frac{d\zeta}{dy}\,[\psi_q^S(\zeta)]^2
\nonumber\\
&&=\frac{e_q^0\,Mm_s}{16\pi^2}\int_{-\infty}^1x dx\,\frac{d\zeta}{dx}\,[\psi_q^S(\zeta)]^2
\nonumber\\
&&=-e_q^0\int_{-\infty}^1x dx\,\frac{df_q(x)}{dx}\,
\nonumber\\
&&=e^0_q\int_{-\infty}^1dx\,f_q(x)
\label{eq:norm3}
\eea
in agreement with the results presented in Ref.~\cite{Gross:2006fg}.  The forms (\ref{eq:norm2}) and (\ref{eq:norm3}) are equivalent, alternative forms of the normalization condition.

In this model, $f_q$ is the valence quark distribution, and hence we {\it require\/} its first moment to be unity
\bea
1=\int_0^1 dx f_q(x) \label{eq:normB}
\eea
(where, by our convention, the factor of 2 that accompanies the first moment of the $u$ quark distribution in the proton is contained in the formulae (\ref{eq:335aa}) and not in the moment).  This will set the scale of the wave function.  The CST normalization condition (\ref{eq:normCST}) will then determine the renormalization of the quark charge at $Q^2=0$.  This is a different procedure than we used in Ref.\ \cite{Gross:2006fg}, and more in keeping with QCD, which fixes the quark charges at $Q^2\to\infty$.

\subsection{Where is the glue?} \label{sec:IIG}

Gluons are known to make a substantial contribution to the nucleon momentum.  This shows up in momentum sum rule.  Using our normalization, the proton momentum sum rule is
\bea
1=2\int_0^1dx\,x f_u(x)+\int_0^1 dx\,x f_d(x) +N_g
\eea
where $N_g\simeq0.5$ is the contribution from gluons.   This sum rule cannot be derived within the context of the CST model description of DIS scattering.  
Instead, we have the charge normalization condition (\ref{eq:norm3}).  
If the $u$ and $d$ distributions are identical (for purposes of discussion), and $N_g=0.5$, these two sum rules are
\bea
&&0.167=\int_0^1 dx \,x\,f_q^S(x)
\nonumber\\
&&1=e_q^0\int_{-\infty}^1 dx\,f_q^S(x)\, . \label{eq:normBB}
\eea
In CST, the first sum rule is fixed phenomenologically
(by fitting the theoretical $f_q$ to experiment) and then the dressed charge $e_q^0$ is determined from the second.

To illustrate the idea, choose the oversimplified distribution 
\bea
\widetilde f_q^S(y)=\delta(y-y_0)+\delta\left(\frac1y-y_0\right)
\eea
where $\widetilde f_q^S(y)$ satisfies the symmetry condition (\ref{eq:ysymm}) (as it must), and $y_0$ is a parameter.  Then,
the normalization condition (\ref{eq:normB}) is automatically satisfied, and the conditions (\ref{eq:normBB}) become
\bea
\int_0^1x\,dx f_q^S(x)&=&1-y_0=0.167
\nonumber\\
1&=&e_q^0\left(1+\frac1{y_0^2}\right)\, .
\eea
 These equations give $e_q^0=0.41$, not too far from the values obtained in the following sections.
 
 It remains to be shown in detail how the gluon contributions give rise to the modification of the quark charge, and to generalize this discussion to show how the angular momentum contributions of {\it constituent\/} quarks, evaluated in this paper, can be compared to the angular momentum contributions from bare quarks plus gluons that would be obtained from a light-front model.  This is beyond the scope of this paper and a subject for future work.

\section{First observations}\label{sec:III}

In this section we first present the data for the unpolarized structure functions $f_q$ and the polarized $g_1$, and then discuss the proton and neutron spin puzzles.  The detailed fits will be discussed in Sec.\ \ref{sec:IV}.  

\subsection{Data}

The individual $u$ and $d$ quark distributions $f_q$ and $g_1^q$ have been extracted from global fits to data.  These fits use the QCD evolution equations to relate the data at higher $Q^2$ to phenomenological starting distributions defined at $Q^2=1$ GeV$^2$.  In fitting our model for  $f_q$ and $g_1^q$ to these starting distributions  we assume that it is appropriate to compare the DIS limit of our model  with the data at $Q^2=1$ GeV$^2$,  assuming that the behavior at higher $Q^2$ can be predicted by QCD but not by the model.  Some investigators have chosen to extrapolate the QCD predictions to (much) lower $Q^2$ and fit their models there.  We do not do so for two reasons; (i) we are doubtful that the   QCD evolution equations are reliable below $Q^2=1$ (they are based on perturbative QCD), and (ii) the assumption that our model has reached its asymptotic limit at the scale of the nucleon mass is generous.  The choice of  $Q^2=1$ GeV$^2$ is somewhat arbitrary, and at best a compromise required by trying to fit a round peg into a square hole.

For the fits to $f_q$ we use the global fits of Martin, Roberts, Stirling and Thorne (MRST02) \cite{Martin:2002dr}: 
\bea
x\,f^{\rm exp}_u(x)\to xu_V(x)/2=&&\,{\bf 0.130}\, x^{0.31}(1-x)^{3.50}
\nonumber\\
&&\times(1+3.83 \sqrt{x}+37.65x)
\nonumber\\
x\,f^{\rm exp}_d(x)\to xd_V(x)\quad=&&\,{\bf 0.061322}\, x^{0.35}(1-x)^{4.03}
\nonumber\\
&&\times(1+49.05\sqrt{x}+8.65x)\, ,\qquad \label{eq:u&d}
\eea
where we have divided their $u$ quark distribution by 2 because both of our $f_q(x)$ distributions are normalized to unity
\bea
\int_0^1 dx\,f^{\rm exp}_q(x)=1\, . \label{eq:3.21a}
\eea
[In order to get this valence quark normalization condition accurately, we we rescaled the MRST02 from 0.131 to 0.130 (for $u$ quarks) and from 0.061 to 0.61332 (for $d$ quarks).  These rescaled numbers are shown in bold in Eq.\ (\ref{eq:u&d}).]
The model does not describe sea quarks, so it is appropriate to use the valence quark distributions; the description of sea quarks is a subject for future study.

\begin{figure}
\leftline{
\mbox{
\includegraphics[width=2.8in] {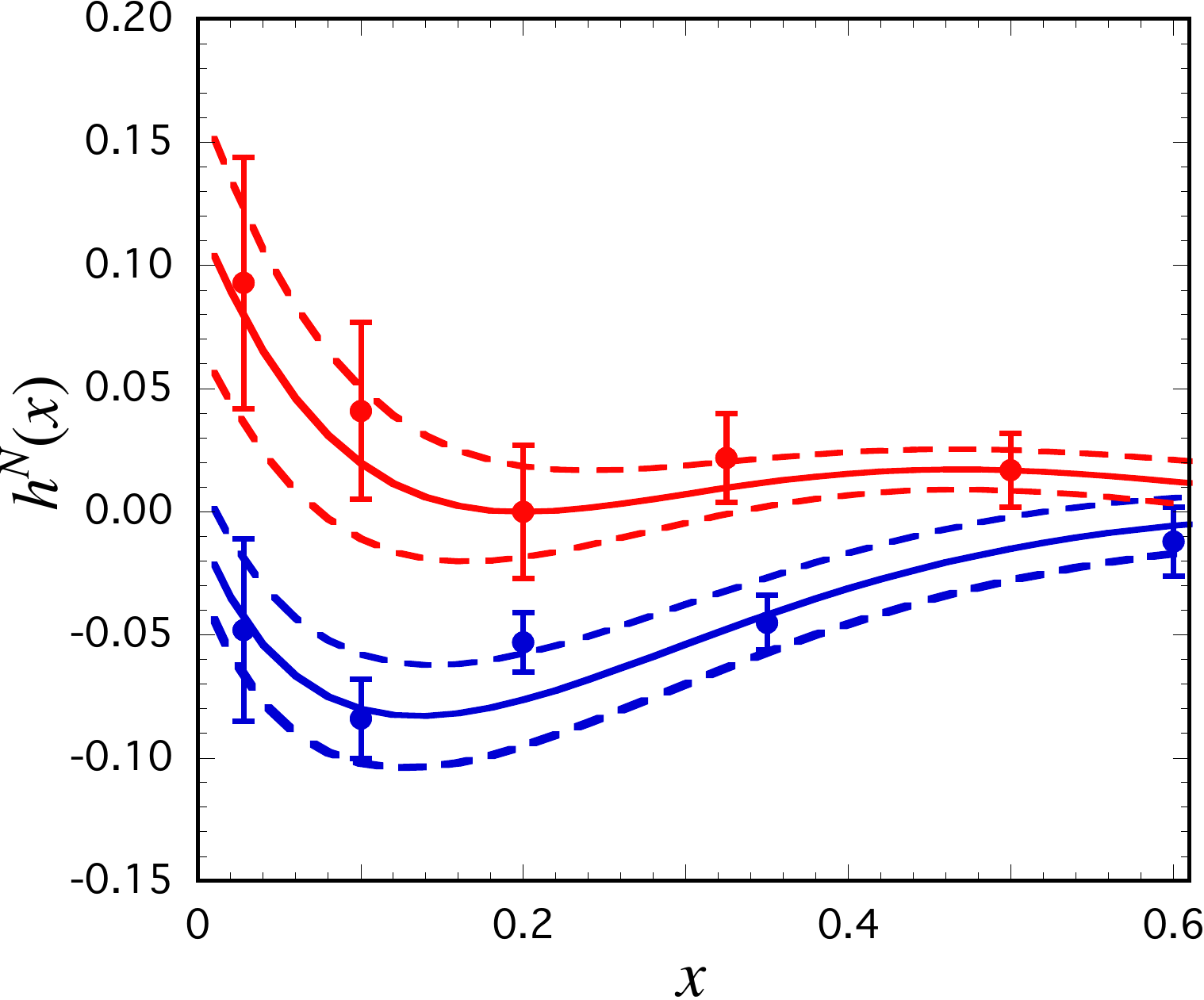}}} 
\caption{\footnotesize{(Color on line) The data for $h^N(x)$ and the empirical fits of Eq.\ (\ref{eq;hfits}).  The dashed lines above and below the solid curves are the error estimates of Eq.\ (\ref{eq;hfits}).  Lower points and curves are the proton; upper are the neutron. }}
\label{fig:2xx}
\end{figure}

\begin{figure*}
\centerline{
\mbox{
\includegraphics[width=6.5in]{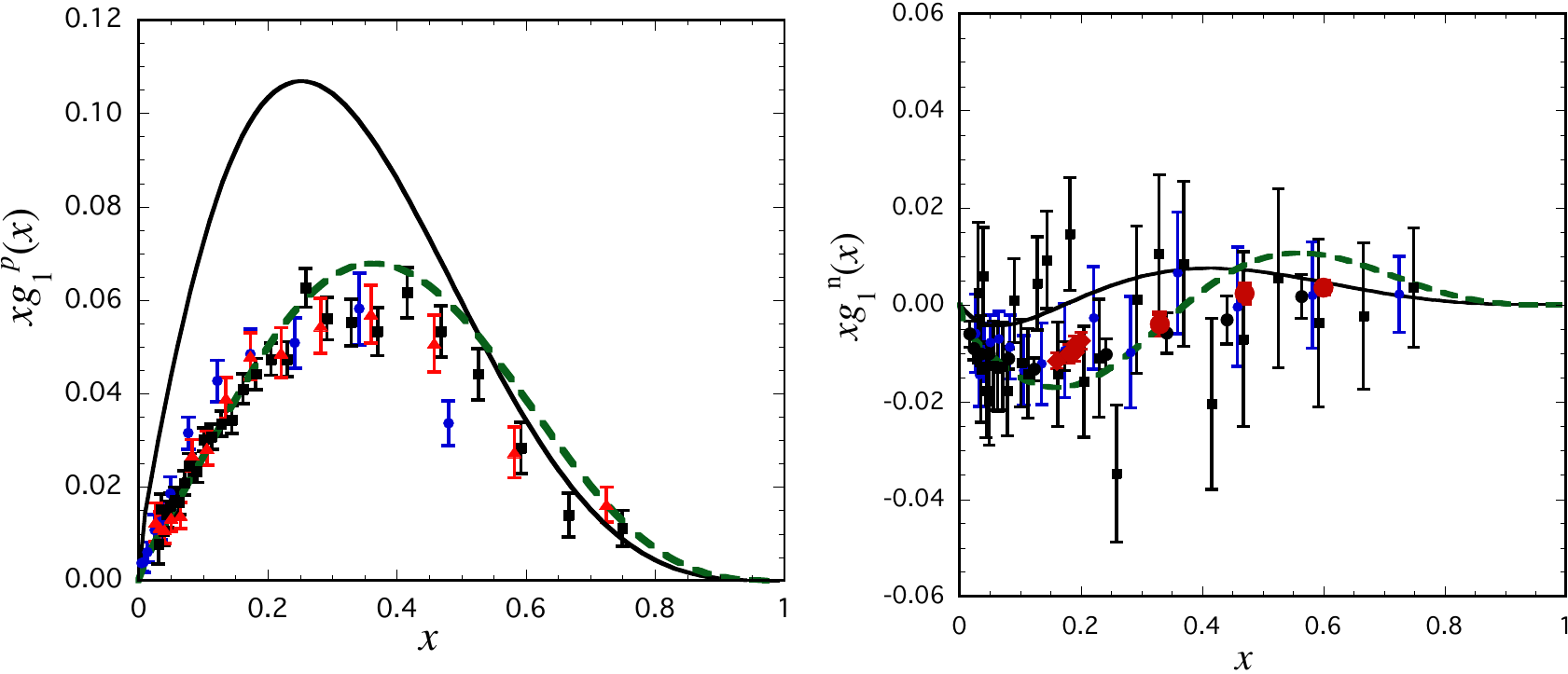} }}
\caption{\footnotesize{(Color on line)  Data for $xg_1$  compared to the predictions (\ref{eq:psf}) (solid line) and the LSS10 fits (for $\delta=0$) (\ref{eq:fitg1}) (dashed line).  
Left panel: proton with data from SMC (circles, Ref.\ \cite{Adeva:1998vv}), SLAC-E143 (squares, Ref.\ \cite{Abe:1998wq}), and HERMES (triangles, Ref.\ \cite{Airapetian:2007mh}); right panel: neutron with data from SLAC-143 (squares, Ref.\ \cite{Abe:1998wq}), HERMES (small circles at low $x$, Ref.\ \cite{Airapetian:2007mh}), SLAC-E154 (medium circles, Ref.\ \cite{Abe:1997qk}), JLab-Hall A (large circles, Ref.\ \cite{Zheng:2004ce}), and JLab-Kramer (diamonds, Ref.\ \cite{Kramer:2005qe}) .  
}}
\label{fig:1b}
\end{figure*}

To represent the data for $g_1^q$, we use the global fits  of Leader, Sidorov, and Stamenov (LSS10) \cite{Leader:2010rb}.  They express these distributions as a sum of the leading twist component, $\Delta q$, and a higher twist component proportional to $h^q$ 
\bea
g_1^q(x)=\Delta q(x)+\frac{h^q(x)}{Q^2}
\eea
At $Q^2=1$, the leading twist contributions determined by LSS10  are
\bea
x\Delta u(x)=&&0.548\, x^{0.782}(1-x)^{3.335} 
\nonumber\\&&\times
(1-1.779\sqrt{x}+10.2\, x)
\nonumber\\
x\Delta d(x)=&&-0.394\, x^{0.547}(1-x)^{4.056} 
(1+6.758\, x)\, .\qquad \label{eq:LSSfit}
\eea
However, the higher twist corrections are not negligible at $Q^2=1$.  Extracted values of $h^N$ for the proton and neutron are reported  in Table III of Ref.\ \cite{Leader:2010rb}.  For convenience these were fit by the smooth functions;
\bea
h_\pm^p(x)&=&-0.82\, x^{0.782} (1-x)^5[1\pm \delta_p(x)]
\nonumber\\
h_\pm^n(x)&=&3 (1-x)^4(0.2-x)^2[1\pm\delta_n(x)]\, , \label{eq;hfits}
\eea
where the errors were approximated by
\bea
\delta_p(x)&=&\frac{0.05}{(1-x)^4}+\frac{0.04}{x^{0.7}}
\nonumber\\
 \delta_n(x)&=&\frac{0.015}{(0.2-x)^2}+\frac{0.04}{(1-x)^3}\, .
\eea
The quality of these fits and the errors in the $h$'s are shown in Fig.\ \ref{fig:2xx}.  Separating these into $u$ and $d$ quark contributions and adding them to (\ref{eq:LSSfit}), the total empirical distributions (adding errors) at $Q^2=1$  become
\bea
x\,g^{\rm exp\pm}_{1u}(x)&=&x\Delta u(x)+\sfrac35x[4h_\pm^p(x)-h_\pm^n(x)]
\nonumber\\
x\, g^{\rm exp\pm}_{1d}(x)&=&x\Delta d(x) +\sfrac65x[4h_\pm^n(x)-h_\pm^p(x)]\, .  \label{eq:fitg1}
\eea
As Fig.\ \ref{fig:2xx} shows, the error in these functions is considerable.

Forming the combinations for proton and neutron, and integrating over $x$  gives
\bea
\Gamma_1^p&=&\int_0^1 dx\,g^{\rm exp\pm}_{1p}(x)=0.128\pm 0.013
\nonumber\\
\Gamma_1^n&=&\int_0^1 dx\,g^{\rm exp\pm}_{1n}(x)=-0.042\pm 0.013\, .\label{eq:exptgamma1} 
\eea
These values are compatible with recent measurements reported at a number of experimental facilities \cite{Kuhn:2008sy,Chen:2010qc}, and also agree, within errors, with the result for the proton reported by Jaffe and Manohar  \cite{Jaffe:1989jz}.

For later use, we record the experimental values of $\Gamma_1$ for the separate $u$ and $d$ distributions obtained from the experimental results using the expansion (\ref{eq:polex}):
\bea
\Gamma_1^u&=&\sfrac35\Big(4\Gamma_1^p-\Gamma_1^n\Big)=0.333\pm 0.039
\nonumber\\
\Gamma_1^d&=&\sfrac65\Big(4\Gamma_1^n-\Gamma_1^p\Big)=-0.355\pm 0.080\, ,\label{eq:exptgamma} 
\eea
where the errors are estimates obtained by integrating Eq.\ (\ref{eq:fitg1}).

\subsection{Proton and neutron spin puzzles}

Note that, if the nucleon has no P or D-state components, the polarized spin structure functions $g_1^n$ are uniquely predicted.  From Eq.\ (\ref{eq:335aa}) we obtain 
\bea
g^p_1(x) =&&\sfrac{11}{30}\,f_p -\sfrac2{15} f_n  = \sfrac8{27}\,f_u-\sfrac1{54}\,f_d 
\nonumber\\
g^n_1(x) =&&\sfrac{2}{15}\,f_p -\sfrac3{15} f_n=\sfrac2{27}\,f_u-\sfrac2{27}\,f_d
 \label{eq:psf}
\eea
Recalling the normalization (\ref{eq:3.21a}), the moments predicted by (\ref{eq:psf}) are
\bea
\Gamma_1^p&=&\int_0^1 dx\, g^p_1(x)=\sfrac5{18}=0.278
\nonumber\\
\Gamma_1^n&=&\int_0^1 dx\, g^n_1(x)=0
\eea
These predictions are to be compared with the experimental values (\ref{eq:exptgamma1}).  As mentioned in the Introduction, the inability to correctly predict $\Gamma_1^p$ has been referred to as the proton spin puzzle, and we see here that our model without orbital angular momentum components cannot reproduce the experimental results.   From our point of view, the neutron spin puzzle is also interesting.

The extent to which the prediction (\ref{eq:psf})  for $g_1^p$ strongly disagrees with the data is shown in the left panel of Fig.~\ref{fig:1b}; the prediction for $g_1^n$ shown in the right panel is in better agreement.   This figure is included here only as an illustration of the spin problem.  The data shown  are only  a subset of all the data that is available and were measured at a variety of $Q^2$, which, because of  the effects of QCD evolution, should strictly speaking not be placed on the same plot.  The method we used to fit the data, which allows for QCD evolution and includes a larger data set, is discussed in the next section.

\section{Fits to the data} \label{sec:IV}



The fits to the DIS data will be done in three steps.  First, the unpolarized structure functions $f_q(x)$ will be fit using a model with only an S-state component. 

After the S-state component has been fixed, we use the same parameters for the P and D-state components and adjust the strength parameters $n_P$ and $n_D$  to get the correct values of $\Gamma_1^u$ and $\Gamma_1^d$ given in (\ref{eq:exptgamma}).  Once the $n_P$ and $n_D$ have been chosen, we readjust some of the parameters of the wave functions to give a good fit to the shapes.  This last step confirms that our choices of $n_P$ and $n_D$ made in step 2 were acceptable.  However, this procedure is rather crude, and a fit to all of the parameters {\it at once\/} would likely alter our conclusions somewhat.

\subsection{Step 1: fitting the S-state wave functions}

As described above, the S-state $u$ and $d$  quark distributions are are fit to the experimental quark distributions $f_u$ and $f_d$ using  Eq.~(\ref{eq:3.7}).  We choose a simple form for the wave functions, with  parameters adjusted to give a reasonable fit to the data.   

In our previous work \cite{Gross:2006fg} we made the choice
\bea
\psi_q^S(\chi)=\frac{N_S}{m_s[\chi+\beta_1][\chi+\beta_2]}\, ,
\eea
where $N_S$ is a normalization constant, $\beta_1$ and $\beta_2$ are dimensionless range parameters, and $m_s$ is included to make the normalization constant dimensionless.  This form, used in our previous analysis of the nucleon form factors, does not do well fitting the DIS data, and, as discussed in the Introduction, we will adopt a completely different approach in this paper.  

\begin{table}[b]
\begin{minipage}{3.5in}
\begin{center}
\begin{tabular}{lcccccc}
& $\quad\beta_{Sq}\quad$ & $\quad \theta_{Sq}\quad$ & $\quad n_{0Sq}\quad $  & $\quad n_{1Sq}\quad $ & $\quad C^S_q\quad$ & $\qquad e^0_q\qquad$ \\
\hline
$u$ & {\bf 0.9} & {\bf 0.4}$\pi$  & {\bf 0.51} &  3  & 2.197 &0.3545 \\
$d$  & {\bf 1.25} & $\sfrac14 \pi$ &  {\bf 0.49} & {\bf 3.2} &  2.279  &0.3940\\[0.05in]
\hline
\end{tabular}
\end{center}
\caption{Adjustable parameters (in bold) and additional  constants determined by the  normalization conditions for the fits shown in Fig.\ \ref{fig:6}.}
\label{tab:1}
\end{minipage}
\end{table}

A choice that produces a good fit to the DIS data is
\bea
\psi_q^S(\chi)&=&\frac1{m_s N^S_q}\;\frac{\beta_{Sq}\cos \theta_{Sq}+\chi\sin \theta_{Sq}}{\chi^{n_{0Sq}}[\chi+\beta_{Sq}]^{n_{1Sq}-n_{0Sq}}} 
\nonumber\\
&\equiv&(m_sN^S_q)^{-1} \Phi_q^S(\chi,\beta,\theta,n_{0},n_{1})
\label{eq:3.14a}
\eea
where $N^S_q$ is a normalization constant, $\beta_{Sq}$ is the single dimensionless range parameter, $\theta_{Sq}\equiv a_{Sq}\,\pi$ a mixing parameter which allows the phase and/or oscillations of the wave function to be adjusted as needed, $n_{0Sq}$  is a fractional power needed to give the sharp rise in the distributions amplitudes at small $x$, and $n_{1Sq}$ allows for adjustment of the large $x$ behavior of the wave function.   The generic wave function, $\Phi(\chi,\beta,\theta,n_0,n_1)$, defined in (\ref{eq:3.14a}), will also be used to define the P and D-state wave functions below, and we use the notation $\Phi_q^S(\chi, \beta, \cdots)=\Phi(\chi,\beta_{Sq},\cdots)$.    If $a_{Sq}\ne0$ and $n_{1Sq}=3$, Eq.\ (\ref{eq:3.14a}) still goes as $1/\chi^2$ at large $\chi$, ensuring that the form factors calculated from this wave function will go as $1/Q^4$ at large $Q$.  

Taking $r=1$, and using the correspondence presented in Eq.\ (\ref{eq:asywf}), and assuming $\theta_{Sq}\ne0$, the  leading behavior of the distribution amplitudes $f_q$ near $x\to0$ and $x\to1$ should be of the from $x^{\alpha_q}(1-x)^{\gamma_q}$, where
\bea
n_{0Sq}&\sim& \sfrac12-\sfrac14 \alpha_q
\nonumber\\
n_{1Sq}&\sim&\sfrac32+ \sfrac12\gamma_q \, . \label{eq:powers}
\eea
%

\begin{figure}
\centerline{
\mbox{
\includegraphics[width=3.5in]{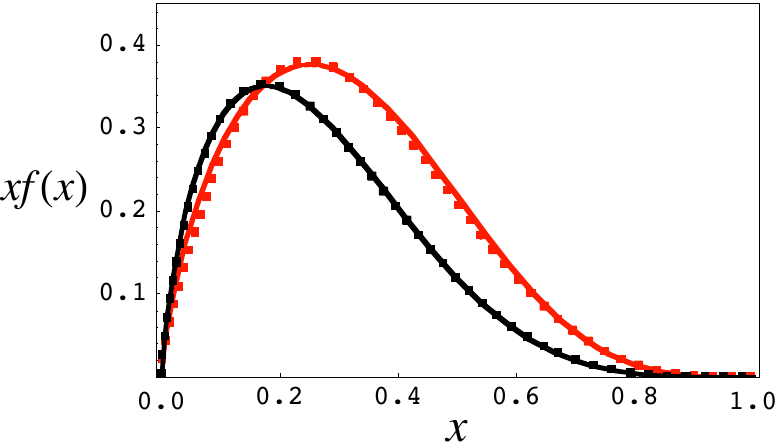} }}    
\caption{\footnotesize{(Color on line) The experimental quark distributions $x f^{\rm exp}_u(x)$ and $x f^{\rm exp}_d(x)$ (dotted lines)  compared to the fits $xf^S_u(x)$ and $xf^S_d(x)$ with the parameters given in Table \ref{tab:1} (solid lines).  All distributions are normalized to unity as in Eq.~(\ref{eq:3.21a}).   The curves that peak at a larger $x$ are  the $u$ distributions, the ones that peak at smaller $x$ are the  $d$ distributions.}}
\label{fig:6}
\end{figure}

\begin{figure}
\centerline{
\mbox{
\includegraphics[width=3in]{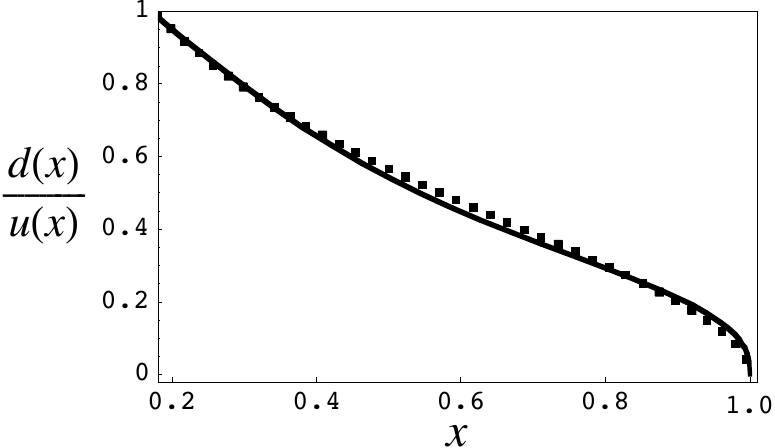} }}
\caption{\footnotesize{Comparison of the fitted to the experimental ratio of the $d$ and $u$ quark distributions, where $q(x)\equiv f_q(x)$. The dotted line represents the
experimental results.}}
\label{fig:6ab}
\end{figure}

For the $u$ quark distribution, we will fix $n_{1Su}=3$, giving a $(1-x)^3$ behavior at large $x$, so this simple model cannot reproduce the fractional power  of $(1-x)^{3.5}$ found in the empirical $u$ quark function (\ref{eq:u&d}).  However, in order to preserve the interesting $d/u$ quark behavior as $x\to 1$, we allow the power $n_{1Sd}>3$, and adjust it to give a good $d/u$  ratio.  

If the wave functions were to reproduce the small $x$ behavior exactly, we would require
\bea
n_{0Sq}\simeq \sfrac12-\sfrac14 \alpha_q=\begin{cases} 0.42 & u\;{\rm quark} \cr 0.41 & d\;{\rm quark}\, .\end{cases} \label{eq:powersa}
\eea
The actual parameters that emerge from our fits are not far from these estimates.

The S-state structure functions reduce to
\bea
 f_q^S(x)&&=\frac1{C_q^S\, r} \int_\zeta^\infty d\chi \,[\Phi_q^S(\chi)]^2\qquad
 \nonumber\\
 &&\equiv (C_q^S)^{-1} F_q^S(x)
\label{eq:fx}
\eea
where, for simplicity, the parameters that enter $\Phi_q^S$ have been suppressed, 
and
\bea
C_q^S=16\pi^2(N^S_q)^2
\eea
To complete the determination of $f_q^S(x)$, $C_q^S$ and $e^0_q<1$ are determined by the wave function normalization condition (\ref{eq:norm3})   [or  (\ref{eq:normCST})] and the normalization of the valence quark distribution (\ref{eq:3.21a}).  The wave function normalization gives $C_q^S$ in terms of $e_q^0$ 
\bea
C_q^S&=&e^0_q\int_{-\infty}^1 dx\, F_q^S(x)\, . \qquad
\label{eq:normC}
\eea
with $e^0_q$ given by the ratio 
\bea
e^0_q= \frac1N\,\int_0^1 dx\, f_q^S(x)
\, ,
\label{eq:eq0}
\eea
with
\bea
N\equiv\int_{-\infty}^1 dx\, f_q^S(x)\, . \label{eq:constN}
\eea


The fit is shown in Fig.~\ref{fig:6} and the parameters are given in Table~\ref{tab:1}.    Note that  wave function is highly singular (but they are normalizable; only if $n_{0Sq}\ge0.75$ can they not be normalized), and even more singular than the estimates (\ref{eq:powersa}).  The ratio of the $d$ to $u$ quark distributions, which vanish at $x\to1$, are shown in Fig.\ \ref{fig:6ab}.  We constrained $n_{1Su}=3$, but we cannot get the correct result for the $d/u$ ratio unless $n_{1Sd}>3$, and choosing the power $n_{1Sd}=3.2$ gives the fit shown in Fig.\ \ref{fig:6ab}.  This means that the asymptotic nucleon form factors will be dominated by the $u$ quark distribution.

\begin{figure*}
\centerline{\mbox{
\includegraphics[height=1.85in]{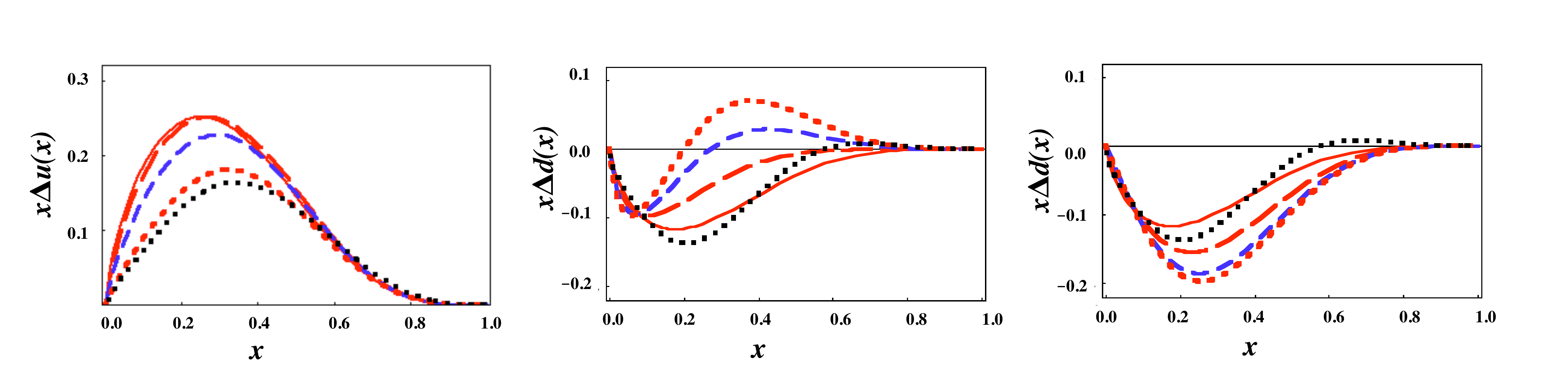}}}
\caption{\footnotesize{(Color on line) Influence of a D-state on the spin polarized structure functions, $x\Delta q(x)$, as a function of $x$      .  In all panels, the dotted line is the experimental fit of LSS10, and the theoretical curves have $|n_D|=0$ (solid), 0.2 (long dashed), 0.4 (medium dashed), and 0.6 (short dashed).  Left panel is $x\Delta u$, middle panel $x\Delta d$ for $n_D>0$, and left panel $x\Delta d$ for $n_D<0$. }}
\label{fig:D}
\end{figure*}


\subsection{Step 2: fixing the P and D-state admixtures}

The strength of the P and D-state admixtures is fixed by fitting the moments of the polarized distributions (\ref{eq:exptgamma1}).  Before doing this, we discuss our choice of the functional form for the P and D-state wave functions, and then look qualitatively at the effect of changing $n_P$ and $n_D$ individually.

\subsubsection{Functional form for the wave functions}

We chose the P and D-state wave functions to have the general form
\bea
\psi_q^P(\chi)&=&[m_s N^P_q M\,K(\chi)]^{-1}\Phi_q^P(\chi,\beta,\theta,n_{0},n_{1}),\qquad
\nonumber\\
\psi_q^D(\chi)&=&[m_s N^D_q M^2K^2(\chi)]^{-1}\Phi_q^D(\chi,\beta,\theta,n_{0},n_{1}),\qquad \label{eq:PDwf}
\eea 
where $q=\{u,d\}$, $N_q^L$ are normalization constants, and the function $K(\chi)$ is just the scaled three-momentum expressed as a function if $\chi$
\bea
K(\chi)=\sqrt{\sfrac14\,r^2\chi(\chi+4)}=\kappa\, , \label{eq:kappachi}
\eea
and the generic functions $\Phi_q^L$ have the same general form as $\Phi_q^S$ but with different parameters associated with each functions [so that, for example, $\Phi^P_q(\beta,\cdots)\to \Phi(\beta_{Pq},\cdots)$].

The factors of $K^{-1}$ multiplying the P-state wave function and  $K^{-2}$ multiplying the the D-state wave function in Eq.\ (\ref{eq:PDwf}) were chosen for convenience; they cancel similar factors in the spin structures multiplying theses scalar wave functions.  Our philosophy is to let the short range behavior be fixed by the structure functions themselves, which are more singular than might be expected.

Using (\ref{eq:z0}) to fix $z_0$, and substituting the general forms 
(\ref{eq:PDwf}), the structure functions that appear in (\ref{kpintb}) and (\ref{kpinta}) consist of terms depending on the squares of the wave functions
\bea
f_q^L(x)=&&\frac1{C_q^L \,r}\int_\zeta^\infty  d\chi\, [\Phi_q^L(\chi)]^2\equiv (C_q^L)^{-1} F_q^L(x)
\nonumber\\
g_q^L(x)=&&\frac1{C_q^L\,r}\int_\zeta^\infty   d\chi\, P_2(z_0)\, [\Phi_q^L(\chi)]^2 \, , \label{eq:412a}
\eea
and interference terms
\bea
d_q(x)=&&\frac1{\sqrt{C_q^S C_q^D}\,r}\int_\zeta^\infty d\chi\,P_2(z_0) \Phi_q^S(\chi)\Phi_q^D(\chi)
\nonumber\\
h^L_q(x)=&&\frac1{\sqrt{C_q^P C_q^L}\,r}\int_\zeta^\infty d\chi\,z_0\,\Phi_q^L(\chi) \Phi_q^P(\chi)
\nonumber\\
h^{L+1}_q(x)=&&\frac1{\sqrt{C_q^P C_q^L}\,r}\int_\zeta^\infty d\chi\frac{K(\zeta)}{4x}(1-z_0^2)\,\Phi_q^L(\chi) \Phi_q^P(\chi)
\nonumber\\
&& \label{eq:412}
\eea
where $\zeta$ was defined in Eq.\ (\ref{eq:3.13}),
\bea
C_q^L=16\pi^2(N_q^L)^2,
\eea
and the $L$'s appropriate to each equation were specified before.  The functions $h^{L+1}_q$ enter into the expressions for $g_2$ only, and will not be needed until the next section.

\begin{table}[b]
\begin{minipage}{3.5in}
\caption{Values of $\Gamma_1^q$ for the four choices of $n_D$ shown in Fig.\ \ref{fig:D}.}
\begin{center}
\begin{tabular}{lccccc}
$|n_D|$& $\qquad 0\quad$ & $\quad 0.2\quad$ & $\quad 0.4\quad $  & $\quad 0.6 \quad $ & expt \\
\hline
$\Gamma_1^u\, (n_D>0)$ &0.667 &0.643 &0.544& 0.367& 0.333 \\
$\Gamma_1^d\, (n_D>0)$  & $-$0.333 & $-$0.293& $-$0.252&$-$0.218& $-$0.355\\
$\Gamma_1^d\, (n_D<0)$  & $-$0.333 & $-$0.369& $-$0.395& $-$0.404& $-$0.355\\
\hline
\end{tabular}
\end{center}
\label{tab:D}
\end{minipage}
\end{table}

The P and D-state wave functions are normalized as in (\ref{eq:normall}), and with the ansatz  (\ref{eq:PDwf}) this leads to a generalization of the condition (\ref{eq:normC})
\bea
C_q^L&=&e^0_q\int_{-\infty}^1 dx\, F_q^L(x).
\eea
However, the valence quark normalization condition needed to  extract $e^0_q$  is now
\bea
1&=&\int_0^1 dx\,f_q(x)
\nonumber\\
&=& \int_0^1 dx\Big[n_S^2f_q^S-2n_Pn_S\,h^0_q+n_P^2\,f_q^P+n_D^2\,f_q^D\Big].\qquad\quad
\eea
Since the individual wave functions are all normalized to the same quantity $(e_q^0)^{-1}$, the constant $N$ defined in (\ref{eq:constN}) is independent of $L$
\bea
N=\int_{-\infty}^1 dx\, f_q^P(x) =\int_{-\infty}^1 dx\, f_q^D(x)\, ,  
\eea
so that  the new result for $e_q^0$ is a generalization of (\ref{eq:eq0})
\bea
e^0_q
&=&n_S^2 \,e^S_q-2n_Pn_S\,h^{SP}_q+n_P^2\,e_q^P+n_D^2\,e_q^D\qquad \label{eq:fulleq0}
\eea
where the new coefficients $e_q^L$ are generalizations of (\ref{eq:eq0})
\bea
e^L_q\equiv \frac1N \int_0^1 dx\, f_q^L(x)
\qquad h^{SP}_q\equiv \frac1N \int_0^1 dx\, h_q^0(x)
\, .\qquad
\label{eq:eqL}
\eea

Next the choice of the coefficients $n_P$ and $n_D$ is discussed.

\subsubsection{Effects of the D-state}

First, consider the results of varying the D-state admixture with the P-state component {\it identically zero\/}. Keeping the parameters for the D-state wave function identical to those for the S-state (Table \ref{tab:1}),  the predictions of Eq.\ (\ref{eq:theoryquark}) reduce to (ignoring the small $g_q^D$ terms here, but not in the calculations and figures discussed below)
\bea
 g_1^u&=&
 \sfrac23 f_u^S-n_D^2\,f_u^D -\sfrac29\,a_{SD}\,d_u
\nonumber\\
g_1^d&=&
-\sfrac13 f^S_d
-\sfrac89 a_{SD}\,d_d  
\, ,\qquad  \label{eq:theoryquarkD}
\eea
where, in this case, $f_q=f_q^S$.  The results for various values of $n_D$ are summarized in Fig.\ \ref{fig:D} and Table \ref{tab:D}.  Note that, when the D-state parameters are identical to the S-state parameters, the normalization of the charge is unchanged, i.e. $e_q^0=e_q^S$.  Since the unpolarized distributions $f_q(x)$ are identical to the S-state distributions shown in Fig.\ \ref{fig:6}, Fig.\ \ref{fig:D} does not show $f_q$. 

\begin{figure*}
\centerline{\mbox{
\includegraphics[height=3.5in]{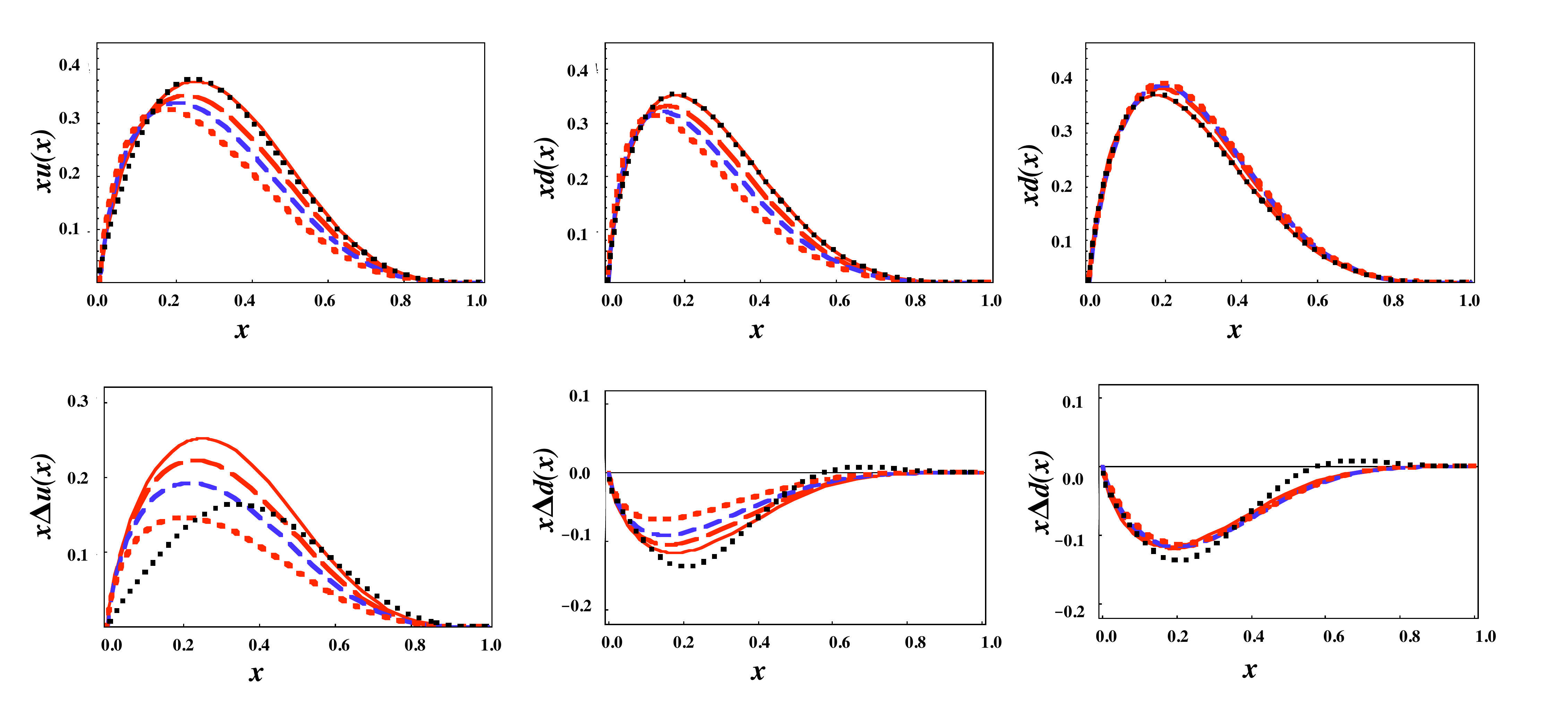}}}
\caption{\footnotesize{(Color on line) Top row: unpolarized structure functions $xq=xf_q$; second row: spin polarized structure functions, $x\Delta q(x)$, all as a function of $x$.  In all panels, the dotted line is the experimental fit of MRST02 or LSS10, and the theoretical curves have $|n_P|=0$ (solid), 0.2 (long dashed), 0.3 (medium dashed), and 0.4 (short dashed).  Left panels are $xu$ and $x\Delta u$, middle panels are $xd$ and $x\Delta d$ for $n_P>0$, and left panels $xd$ and $x\Delta d$ for $n_P<0$. }}
\label{fig:P}
\end{figure*}

As the left panel of Fig.\ \ref{fig:D} and the first line of Table \ref{tab:D} show, a D-state component of about 36\% will fit the $u$ quark polarization and $\Gamma_1^u$ very well.  It also shows that the sign of $n_D$ must be positive in order for the interference term, $d_u$, to bring the predicted $g_1^u$ down close to the observed shape.  [A detailed study of Eq.\ (\ref{eq:theoryquarkD}) reveals that a large negative $n_D$ close to unity is also mathematically possible, but this case is rejected on physical grounds.]   Unfortunately, the $d$ quark distribution, fit well by $n_D=0$, is destroyed by increasing $|n_D|$ to 0.6.  Since the phase of the $d$ quark D-state wave function can be chosen independently of the phase of the $u$ quark wave function,  both possible choices of sign for the $d$ quark distribution are shown, but neither will correct the problem.  The principal source of the difficulty is the relative size of the SD interference them, which is four times larger for $d$ quarks than for $u$ quarks, and is not compensated by any positive term of the form $n_D^2f_d^D$ [the term proportional to $n_D^2\,g_d^D$, neglected in the estimate (\ref{eq:theoryquarkD}), is positive, but much to small to have the desired effect]. 

However, as will be shown, a reasonable fit can be obtained by adding a small P-state contribution and adjusting the parameters of the D-state distributions.

\begin{table}[b]
\begin{minipage}{3.5in}
\caption{Values of $\Gamma_1^q$ and $e_q^0$ for the four choices of $n_P$ shown in Fig.\ \ref{fig:P}.}
\begin{center}
\begin{tabular}{lccccc}
$|n_P|$& $\qquad 0\quad$ & $\quad 0.2\quad$ & $\quad 0.3\quad $  & $\quad 0.4 \quad $ & expt \\
\hline
$\Gamma_1^u\, (n_P>0)$ &0.667 &0.627 &0.563& 0.448& 0.333 \\
$e_u^0\, (n_P>0)$ &0.355 &0.269 &0.229& 0.194 & -- \\
$\Gamma_1^d\, (n_P>0)$  & $-$0.333 & $-$0.313& $-$0.280&$-$0.223& $-$0.355\\
$e_d^0\, (n_P>0)$ &0.394 &0.303 &0.261& 0.224& -- \\
$\Gamma_1^d\, (n_P<0)$  & $-$0.333 & $-$0.321& $-$0.307& $-$0.289& $-$0.355\\
$e_d^0\, (n_P<0)$ &0.394 &0.485 &0.527& 0.564& -- \\
\hline
\end{tabular}
\end{center}
\label{tab:P}
\end{minipage}
\end{table}

\subsubsection{Effects of the P-state}

Now consider the results of varying the P-state admixture with the D-state component {\it identically zero\/}.  Keeping the parameters for the P-state wave function identical to those for the S-state (Table \ref{tab:1}),  the predictions of Eqs.\  (\ref{eq:fq}) and (\ref{eq:theoryquark}) reduce  to (ignoring the small $g_q^P$ terms here, but not in the calculations and figures discussed below)
\bea
f_q&=&f_q^S-2n_Sn_P h_q^0
\nonumber\\
 g_1^u&=&
 \sfrac23 f^S_u-\sfrac43 n_Sn_Ph^0_u-\sfrac89\,n_P^2\,f_u^P
\nonumber\\
g_1^d&=&
-\sfrac13 f^S_d+\sfrac23 n_Sn_P h_d^0+\sfrac49\,n_P^2\,f_d^P 
\, ,\qquad  \label{eq:theoryquarkP}
\eea
where the SP interference term included in $f_q$ has been separated out explicity, and, following Eq.\ (\ref{eq:fulleq0}), the renormalized charge reduces to
\bea
e_q^0=e_q^S-2n_Sn_P h_q^{SP}\, .
\eea

\begin{table}
\begin{center}
\begin{tabular}{lcccc}
solution & $\quad n_P(u)\quad$ & $\quad n_D(u)\quad$  & $\qquad n_P(d)\quad$    & $\quad n_D(d) \quad$    \\
\hline
$1$ & 0.43 &0.18 & $-$0.43   & $-$0.18  \\
$2$  &  0.08 & 0.59 & 0.08   & $-$0.59 \\
\hline
\end{tabular}
\end{center}
\caption{Values of $n_P(q)$ and $n_D(q)$ for the two solutions studied in this paper.  When the sign of $n_L(u)\ne n_L(d)$ it means that the phase of the wave function differs from that of the S-state. }
\label{tab:PDns}
\end{table}

\begin{table}[b]
\begin{center}
\begin{tabular}{lcccc}
case & $\qquad\Gamma_1^u\qquad$ & $\quad e_u^0\quad$  & $\qquad \Gamma_1^d\qquad$    & $\quad e_d^0 \quad$    \\
\hline
$1_0$ & 0.332 &0.187 & $-$0.356   & 0.571  \\
$1_A$ &0.298 & 0.170 & $-$0.333   &0.567 \\
$2_0$  &  0.334 & 0.326 & $-$0.356   & 0.364 \\
$2_A$ &0.334  & 0.326 &$-$0.294 & 0.364  \\
expt & $\quad$0.333$\pm 0.039\quad$ & -- &  $\quad-$0.355 $\pm 0.08\quad$  & --\\
\hline
\end{tabular}
\end{center}
\caption{Values of $\Gamma_1^q$ and $e_q^0$ for the cases shown in Figs.\ \ref{fig:uPD-1} -- \ref{fig:dPD-2}.  Cases $1_0$ and $2_0$ are solutions 1 and 2 with P and D-state  parameters identical to the S-state parameters of Table \ref{tab:1}; cases $1_A$ and $2_A$ have the parameters given in Table \ref{tab:PDpar}. }
\label{tab:PD}
\end{table}

\begin{table}[b]
\begin{minipage}{3.5in}
\begin{center}
\begin{tabular}{lcccccc}
$q$ & $L$& $\quad\beta_{Lq}\quad$ & $\quad \theta_{Lq}\quad$ & $\quad n_{0Lq}\quad $  & $\quad n_{1Lq}\quad $ & $\quad C^L_q\quad$  \\
\hline
$u_1$ &0 & {\bf 0.7} & {\bf 0.45}$\pi$  & {\bf 0.45} &  3  & 2.525  \\
 &1 &{\bf 0.7} & 0.4$\pi$  & {\bf 0.45} &  3  & 1.473  \\
 &2 &{\bf 0.7} & 0.4$\pi$  & {\bf 0.45} &  3  & 1.473  \\
$d_1$ &0  & {\bf 1.8} & $\sfrac14 \pi$ &  {\bf 0.53} & 3.2 &  5.152  \\
 &1  & {\bf 1.8} & ${\bf -0.2}\pi$ &  {\bf 0.53} & 3.2 &  5.396  \\
 &2  & {\bf 1.8} & ${\bf -0.36} \pi$ &  {\bf 0.53} & 3.2 &  22.153 
 \\
 $u_2$ &0 & 0.9 & 0.4$\pi$  & 0.51 &  3  & 2.197  \\
 &1 &0.9 & 0.4$\pi$  &0.51 &  3  & 2.197  \\
 &2 &0.9 & 0.4$\pi$  & 0.51 &  3  &  2.197  \\
$d_2$ &0  & {\bf 0.5} & ${\bf 0.45}\pi$ &  {\bf 0.41} & 3.2 &  1.082  \\
 &1  & {\bf 0.5} & $\sfrac14\pi$ &  {\bf 0.41} & 3.2 &  0.234  \\
 &2  & {\bf 0.5} & ${\bf -0.27} \pi$ &  {\bf 0.41} & 3.2 &  0.497  \\[0.05in]
\hline
\end{tabular}
\end{center}
\caption{Parameters for the cases $1_A$ (denoted by $u_1$ and $d_1$) and $2_A$ (denoted by $u_2$ and $d_2$); those that were adjusted are in bold.  The fits are shown in  Figs.\ \ref{fig:uPD-1} -- \ref{fig:dPD-2}. }
\label{tab:PDpar}
\end{minipage}
\end{table}

The results for various values of $n_P$ are summarized in Fig.\ \ref{fig:P} and Table \ref{tab:P}.  Now that $e_q^0$ and the unpolarized structure functions depend on $n_P$, these quantities are also shown in the table and the figure. 

The equations and figure show that the most efficient way to reduce the polarized $u$ quark distribution to the correct size is to require $n_P>0$ for $u$ quarks, but as the phase of the $d$ quark wave function is independent of the choice for the $u$ quark, we show both cases.  The choice of $n_P=-0.4$ for the $d$ quark preserves the shape of both of the $d$ quark structure functions.  We therefore expect a solution in the region with $n_P>0$ for $u$ and $n_P<0$ for $d$ quarks.

\begin{figure*}
\centerline{
\includegraphics[width=5.3in]{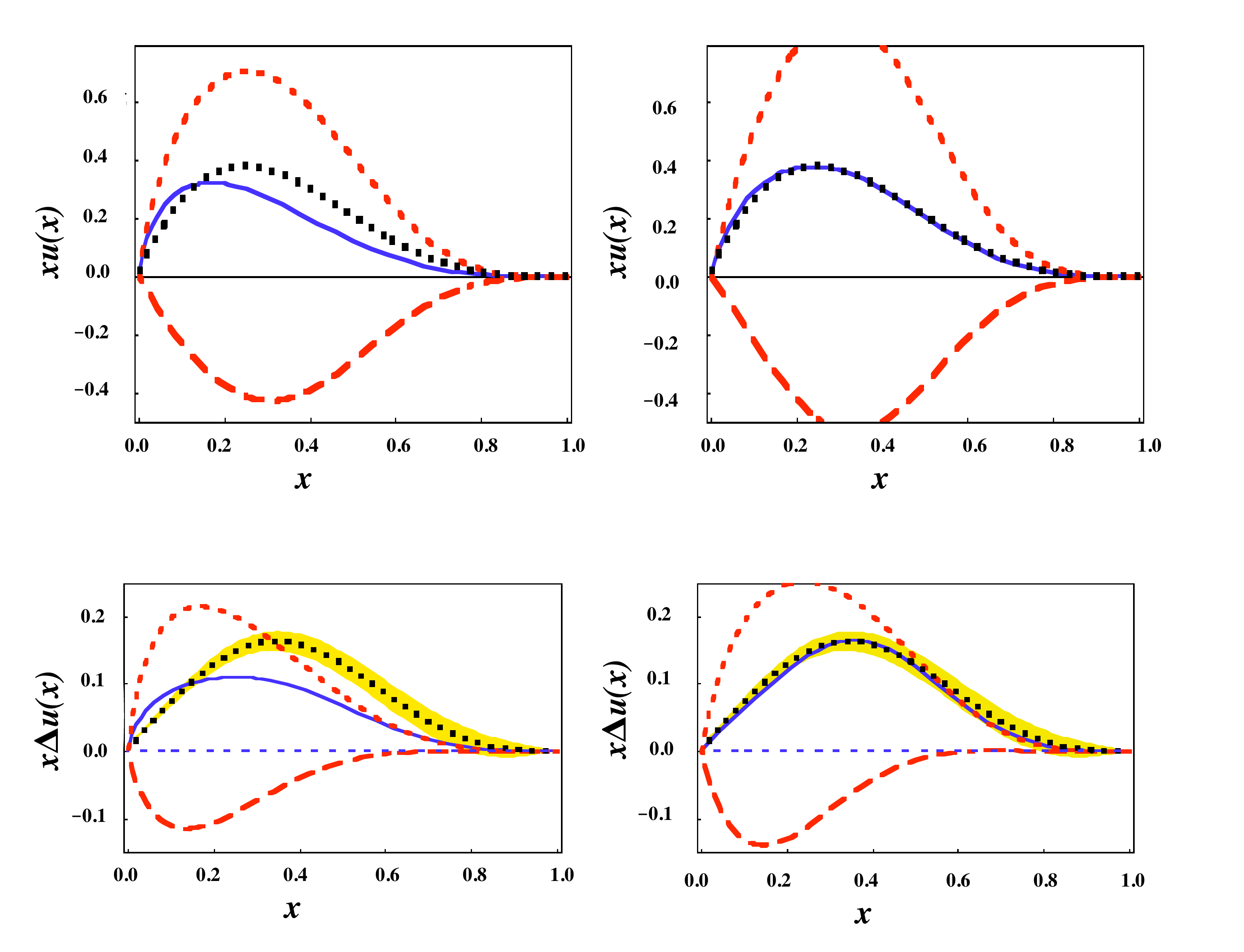}}
\vspace{-0.1in}
\caption{\footnotesize{(Color on line) Distributions for the $u$ quark, as a function of $x$, for solution 1 with $n_P=0.43$ and $n_D=0.18$.  Top row: unpolarized structure functions $xu=xf_u$; second row: spin polarized structure functions, $x\Delta u(x)$; left column: same parameters for all $L$ (case $1_0$); right column: parameters  readjusted to fit the shapes (case $1_A$). }}
\vspace{-0.1in}
\label{fig:uPD-1}
\end{figure*}

\begin{figure*}
\centerline{
\includegraphics[width=5.3in]{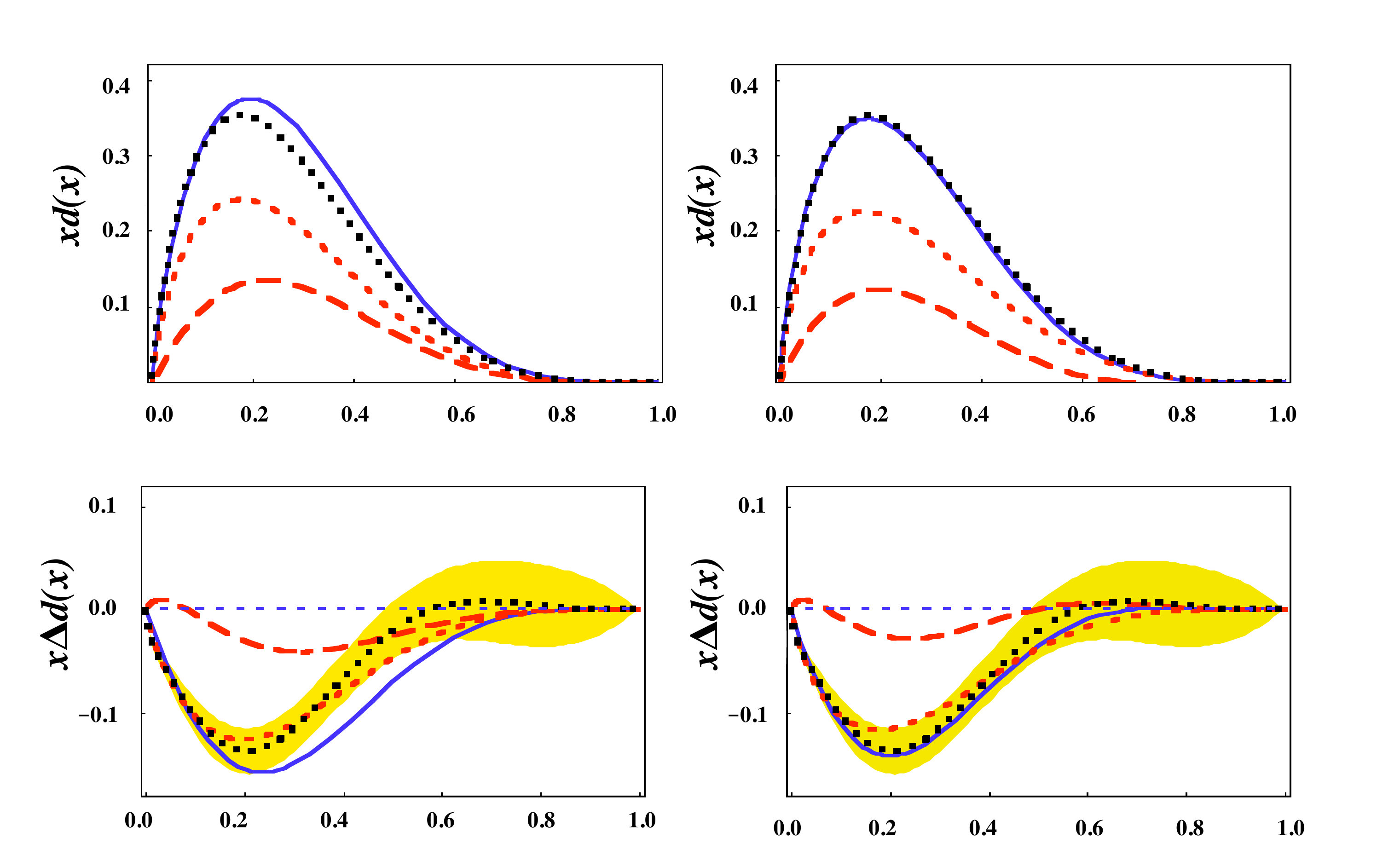}}
\vspace{-0.1in}
\caption{\footnotesize{(Color on line) Distributions for the $d$ quark, as a function of $x$, for solution 1 with $n_P=-0.43$ and $n_D=-0.18$.  See caption to Fig.\ \ref{fig:uPD-1} for more details.}}
\label{fig:dPD-1}
\end{figure*}

\begin{figure*}
\centerline{
\includegraphics[width=5.3in]{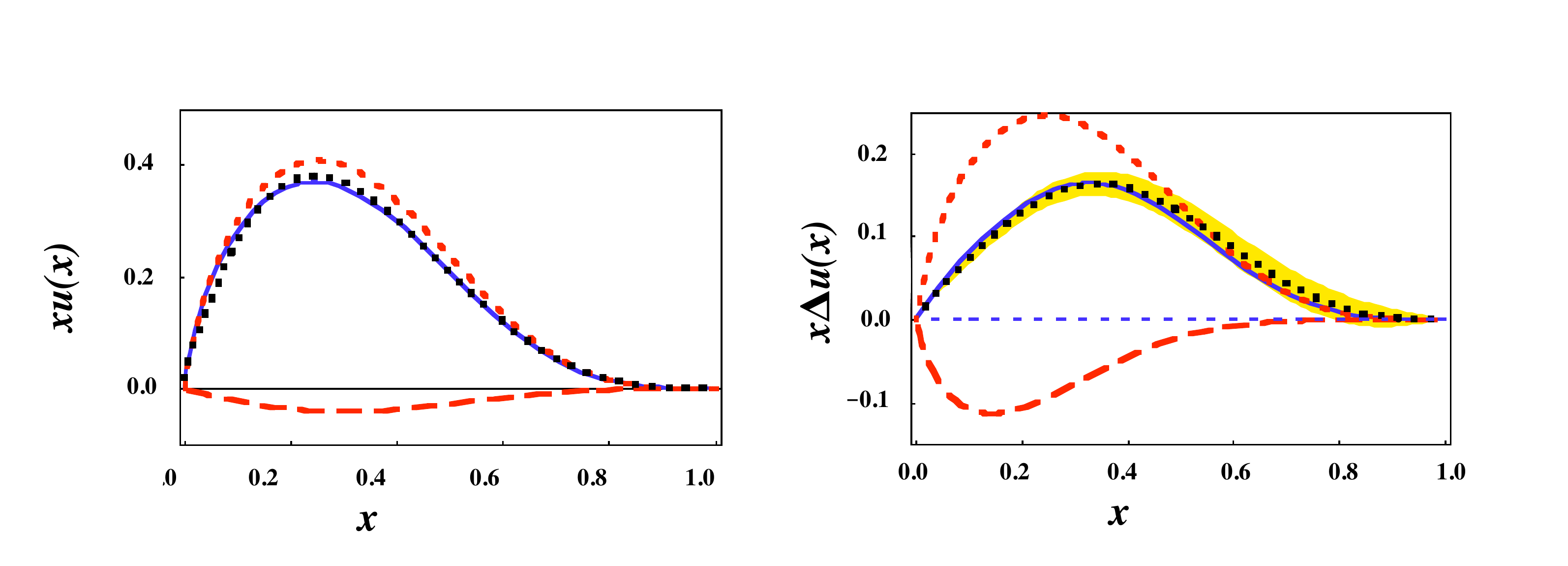}}
\vspace{-0.1in}
\caption{\footnotesize{(Color on line) Distributions for the $u$ quark, as a function of $x$, for solution 2 with $n_P=0.08$ and $n_D=0.59$.  In this case the initial fit was satisfactory, as discussed in the text (case $2_0$ is identical to  $2_A$).}}
\label{fig:uPD-2}
\end{figure*}

\begin{figure*}
\centerline{
\includegraphics[width=5.3in]{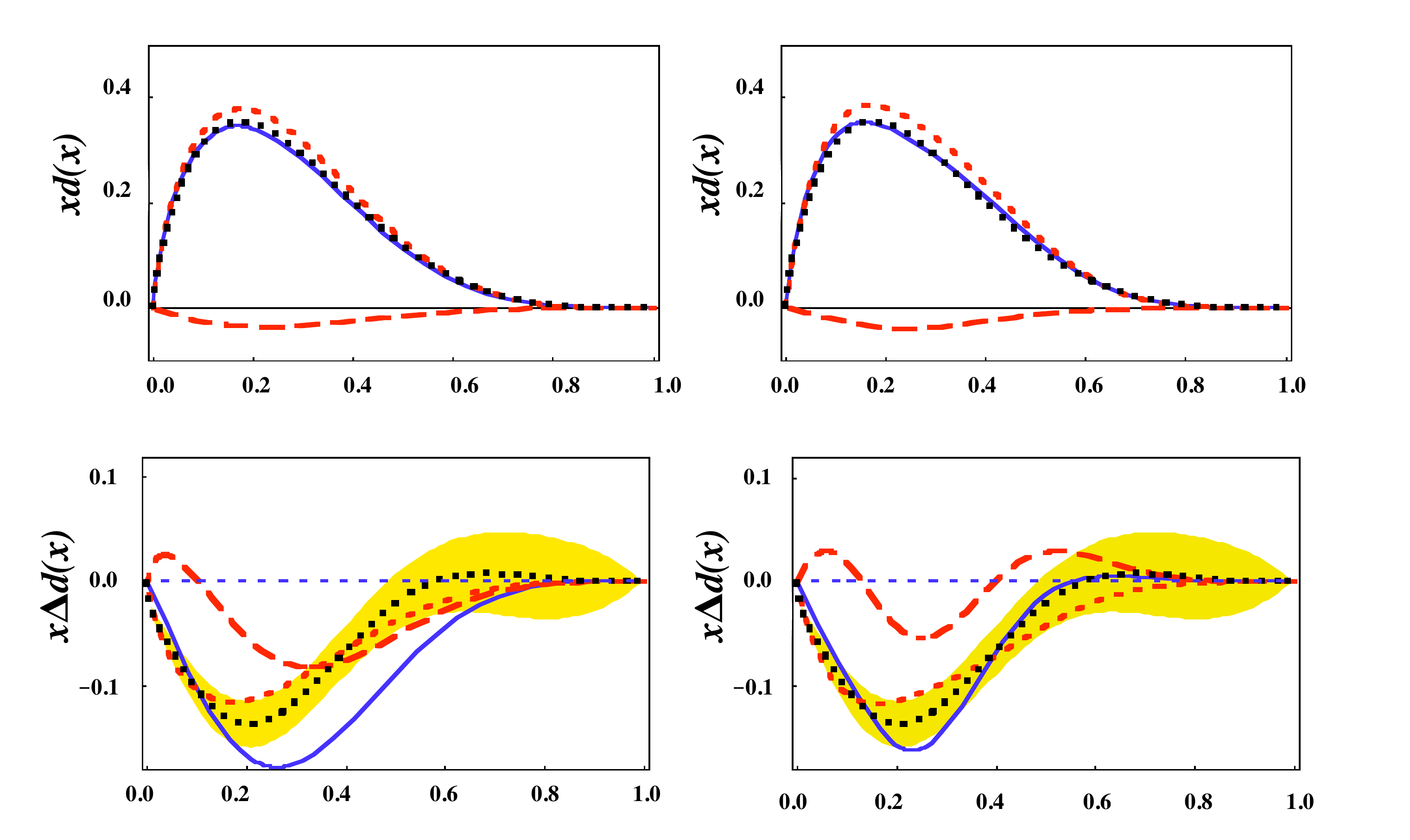}}
\vspace{-0.1in}
\caption{\footnotesize{(Color on line) Distributions for the $d$ quark, as a function of $x$, for solution 1 with $n_P=0.08$ and $n_D=-0.59$ (cases $2_0$ and $2_A$).  See caption to Fig.\ \ref{fig:uPD-1} for more details. }}
\label{fig:dPD-2}
\end{figure*}

\subsubsection{Fixing $n_S$ and $n_D$ from the moments }

An efficient way to estimate the strength of the P and D-state mixing is to adjust the coefficients $n_P$ and $n_D$ to fit the experimental moments $\Gamma_1^q$.  If the parameters of the P and D-state wave functions are set equal to the S-state ones, then the theoretical expression for these moments gives the following equations for $n_P$ and $n_D$
\bea
\Gamma_1^u&=&\sfrac23
-(n_D^2 +\sfrac89 n_P^2)\widetilde F_u-(\sfrac29 a_{SD}-\sfrac89 n_P^2-\sfrac{29}{60}n_D^2) G_u 
\nonumber\\
&&+\sfrac29 a_{PD} H_u =0.333
\nonumber\\
\Gamma_1^d&=&-\sfrac13+\sfrac49 n_P^2)\widetilde F_d-(\sfrac89 a_{SD}+\sfrac49 n_P^2-\sfrac8{15}n_D^2) G_d
\nonumber\\
&&+\sfrac89 a_{PD} H_d=-0.355
\eea
where we require that the solutions to the first equation give positive values of both $n_P$ and $n_D$, but allow solutions with either sign for the second equation, and the coefficients are
\bea
\widetilde F_q&=&\int_0^1 dx\, f_q^{\{0,1,2\}}(x) \qquad  H_q=\int_0^1 dx\, h_q^{\{0,2\}}(x)
\nonumber\\
G_q&=&\int_0^1 dx \,d_q(x)=\int_0^1 dx \,g_q^{\{1,2\}}(x)\, .
\eea
 
With these requirements, there are only two solutions to these equations, summarized in Table \ref{tab:PDns}.  The values of $\Gamma_1^q$ and $e_q^0$ for these two solutions, with the  parameters for all the P and D-state components identical to the S-state, as discussed above, are given in Table \ref{tab:PD}, and the shapes of the quark distributions are shown in Figs.\ \ref{fig:uPD-1} -- \ref{fig:dPD-2}.  Without further changes in the parameters of the individual components, these solutions do not describe the shape of $f_q$ and $g_1^q$, but do give (within round off errors) the correct moments $\Gamma_1^q$.

To complete the fits we move to step 3, adjusting the parameters of the wave functions to give a good description of the shapes.

\begin{figure*}
\centerline{
\mbox{
\includegraphics[width=6.5in]{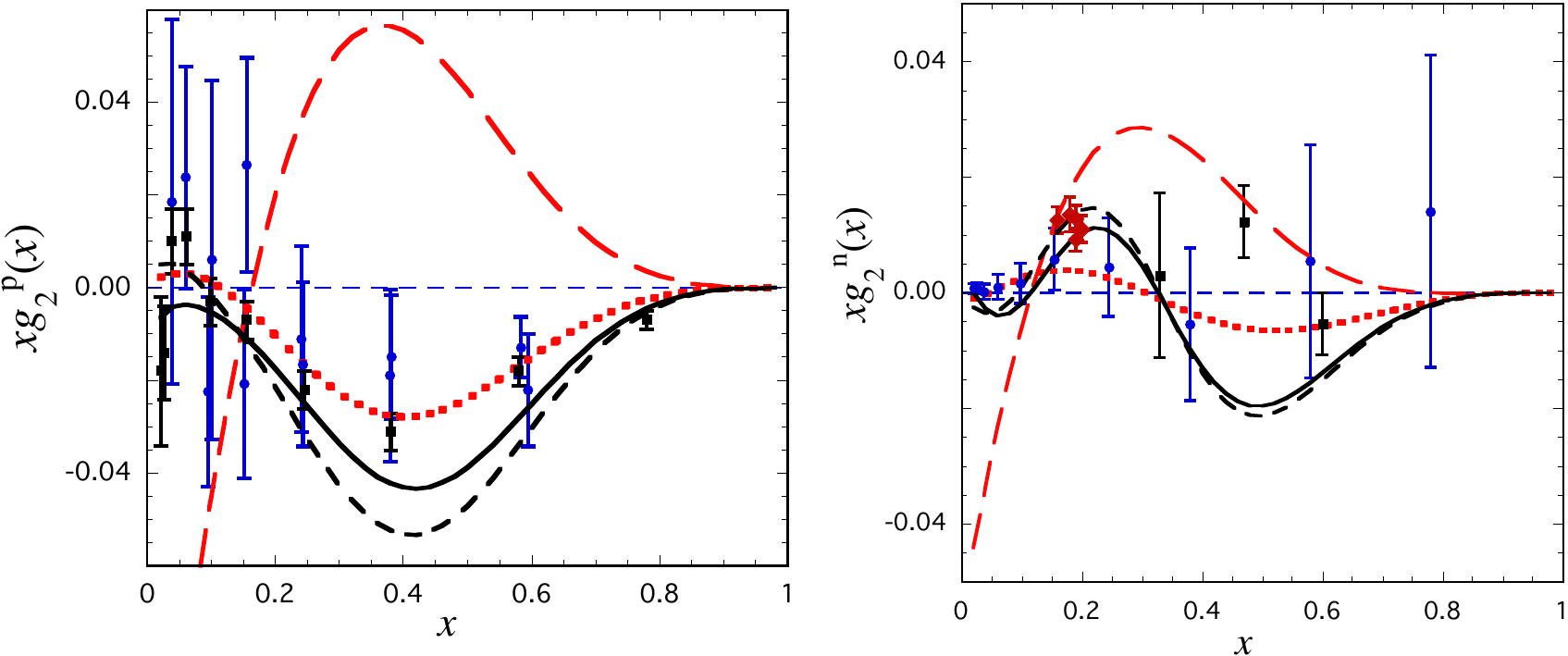} }}
\caption{\footnotesize{(Color on line) Data for $xg_2$ compared to the predictions from Eq.\ (\ref{eq:335aa}) for the two solutions.  In each panel the total result is the long dashed line (solution 1) or solid line (solution 2).  The predictions for $n_P=0$ are the dotted line (solution 1) and short dashed line (solution 2).    Left panel: $xg_2^p$, with data from SLAC-E143 (circles, Ref.\ \cite{Abe:1998wq}) and SLAC-E155 (squares, Ref.\ \cite{Anthony:2002hy}); Right panel: $xg_2^n$, with data from SLAC-E155 (circles, Ref.\ \cite{Anthony:2002hy}),  JLab-HallA (squares, Ref.\ \cite{Zheng:2004ce}), and JLab-Kramer (diamonds, Ref.\ \cite{Kramer:2005qe}). 
}}
\label{fig:1a}
\end{figure*}

\subsection{Step 3: adjusting parameters to fit the shapes}

As a last step, some of the parameters $\beta_{Lq}$, $\theta_{Lq}$, and $n_{0Lq}$ are adjusted to fit the shapes of $f_q$ and $g_1^q$ without destroying the moments $\Gamma_1^q$.  The results are summarized in Table \ref{tab:PD} and Figs.\ \ref{fig:uPD-1} -- \ref{fig:dPD-2}.  In all panels of these figures, the dotted lines are the experimental fits of MRST02 or LSS10 and  the shaded band is our estimate of the experimental errors in $g_1^q$ obtained from the curves in Fig. \ref{fig:2xx} and Eq.\ (\ref{eq;hfits}).  The solid lines show the full theoretical result.  The panels showing $x f_q(x)\equiv x \,q(x)$ also show the individual contributions $x(n_S^2f_q^S+n_P^2f_q^P+n_D^2f_q^D)$ (short dashed line) and $-2n_Sn_Pxh_q^0$ (long dashed line) which add up to the total result $x f_q$.  In the panels showing $x g_1^q(x)\equiv x\Delta q(x)$  the individual results shown are those proportional to $x f_q$ (the first term on the rhs of Eq.\ (\ref{eq:theoryquark}) -- short dashed line) and those proportional to the remaining terms on the rhs of Eq.\ (\ref{eq:theoryquark}), all of which would vanish if $n_P=n_D=0$ (long dashed line).  In the left panels of Figs.\ \ref{fig:uPD-1},  \ref{fig:dPD-1}, and \ref{fig:dPD-2}, the parameters of all of the component wave functions (S, P and D) are given in Table \ref{tab:1}; in the right panels the parameters are given in Table\ \ref{tab:PDpar}.

Note that the final fits and values of $\Gamma_1^q$ all lie within about one standard deviation of the values obtained from the best LSS10 fit.  The parameters space has not be systematically searched: in all cases the values of $\beta$ and $n_{0}$ are equal for each $L$ and $q$, and, to the level of accuracy achievied here, it was not necessary to refit the distributions $u_2$.  The $\theta$ parameters give great flexibility to change the shape of the wave functions, and have been used to significant effect, particularly for the $d$ quark D-state wave functions.  Note that a change in $\theta$ by $\pi$ will merely change the sign of the wave function, leaving its shape  unchanged.  This could be used to fix the $d$ quark values of $n_P$ and $n_D$ to positive numbers.

From our study so far, we can only conclude that the the two solutions 1 and 2 do equally well in describing the structure functions $f_q$ and $g_1^q$.  Significant differences appear when the predictions for $g_2$ are examined.

\section{Predictions for the $g_2$ distributions}\label{sec:g2}

The data  for the structure functions $g_2$ is incomplete and inaccurate but sufficient to allow us to draw some interesting conclusions.

The predictions for $g_2^p$ and $g_2^n$ that follow from the two solutions are shown in Fig.\ \ref{fig:1a}.  The theoretical predictions are given in Eq.\ (\ref{eq:335aa}); note that $g_2$ are the only structure functions that depend explicitly on the interference terms $h^{L+1}_q$, introduced in Eqs.\ (\ref{kpinta}) and (\ref{eq:412}).  

The data for $g_2(x)$ seem to favor the solution 2.  In both panels the total prediction for solution 2 (solid lines) gives a very reasonable approximation of the data, even though the large error bars do not allow us to draw this conclusion too confidently.  But it is clear that the total predictions for solution 1 (long dashed lines) even fail to describe the data qualitatively, particularly for the proton.  (However, it is interesting that both solutions describe the very accurate JLab neutron data equally well.) We conclude that solution 2, with its large D-state, seems to be a more accurate model.  Note that even though this model has a small P-state probability of only about 0.6\%, the P-state does not play a negligible role in the description of the data.  For example, the left panel of Fig.\ \ref{fig:1a} shows that the P-state contribution improves the fit to the data for $x g_2^p(x)$ substantially (compare the short dashed line with the solid line).

Another interesting observation is that the theoretical formulae, Eq.\ (\ref{eq:theoryquark}), show no sign satisfying or even approximating the interesting WW relation \cite{Wandzura:1977qf}
\bea
g_2(x)=-g_1(x)+\int_x^1 \frac{dx'}{x'} g_1(x').
\eea
Numerical studies of the inequalities (for $n\ge0$)
\bea
\int_0^1dx\, x^n \Big[\sfrac{n}{n+1} g_1(x)+g_2(x)\Big]\ll \int_0^1 dx \,x^n  g_1(x)\qquad
\eea
which lead to the WW relation, also show no sign of being satisfied by our model.  Most striking, perhaps, is that the Burkhardt-Cottingham sum rule
\bea
\int_0^1 dx\, g_2(x)=0 
\eea
is satisfied only if $n_D=n_P=0$, but, because of the presence of the structure functions $h^{L+1}_q$ singular as $x\to0$,  diverges otherwise.  Perhaps sea quark contributions, neglected here, are essential for a full understanding of these realtions.  Some of these mysteries may be clarified by future study.

\section{Summary and Conclusions} \label{sec:V}


This paper uses ideas from the Covariant Spectator Theory to develop a covariant constituent quark model of the nucleon that  explains the general size and shape of {\it all\/} of the DIS structure functions of both the proton and the neutron:  $f(x)$ measured in unpolarized scattering and  $g_1(x)$ and $g_2(x)$ measured in polarized scattering.  We find  that a good semi-quantitative description is possible even thought we treat the larger valence quark contributions only, ignoring  the smaller sea quark contributions.  The nucleon wave functions include S, P and D-state contributions, with the sum of their probabilities normalized to unity, and the strength of the P and D-state contributions proportional to the parameters $n_P$ and $n_D$.  These parameters, together with those defining the shape of the wave functions, are adjusted to fit the data for $f(x)$ and $g_1(x)$, and the models defined by these fits are then used to {\it predict\/} $g_2(x)$.

The major effort required to construct the angular momentum components of the wave functions is described in detail in the companion paper \cite{previous} and summarized in Sec.\ \ref{sec:IIB}. In that work we built the general covariant forms for the angular dependence of three-body states with total angular momentum ${L}=1,2$.  Once the wave functions have been parameterized, the calculation of the DIS cross section is straightforward; the details of the evaluation of the traces and the extraction of the DIS limit are discussed in the Appendices.  The parameters of our wave functions for both solutions are given in Table \ref{tab:PDpar}, with the functional form of the S-state described in Eq.\ (\ref{eq:3.14a}) and the P and D-states in Eq.\ (\ref{eq:PDwf}).

We find two solutions that fit the functions $f(x)$ and $g_1(x)$, but only one of these (solution 2) also gives a satisfactory description of $g_2(x)$.  This solution has a large D-state probability of $(0.59)^2\simeq$ 35\% and a small, but important P-state probability of $(0.08)^2 \simeq$ 0.6\%.  The other solution, solution 1, unfavored by the existing data for $g_2(x)$, has a P-state probability of (0.43)$^2\simeq$18\% and a D-state probability of  (0.18)$^2\simeq$3\%, values more in line with previous expectations \cite{Jaffe:1989jz,Isgur:1998yb,Myhrer:2009uq}.    

The observation that the solution with a large D-state exists and fits all of the data is perhaps the most important conclusion of this paper.  It opens the door to new possibilities,  and invites a more serious study of the possible dynamical origin of the D-state components of the nucleon wave function.   Furthermore, the fit requires that the $u$ and $d$ quark  D-state components have a different sign as well as a different shape.

The role that the data for $g_2$ play in this conclusion focuses once again on the importance of accurate measurements of this small quantity.  If the data on $g_2$ were to become sufficiently robust to allow  the separation into $u$ and $d$ quark contributions, it could, from our point of view, be used to decisively fix the size of the P and D components of the wave function of the nucleon.

Of course, many issues remain.  First, we emphasize that our conclusions are sensitive to {\it both\/} the proton {\it and\/} the neutron data, but the extraction of neutron data is not without errors any uncertainties.   And, the modern fits to both  the unpolarized and the polarized data rely on QCD evolution equations to evolve the data to  $Q^2>1$ GeV$^2$ from a fitted phenomenological distribution defined at $Q_0^2=1$ GeV$^2$.  This suggests that $Q_0^2=1$ GeV$^2$ is a good place to match a quark model calculation to QCD, as we have done.  However, the details of our fits would change if we chose to fit the quark model to QCD at a different $Q_0^2$.  Some investigators chose to match the distributions at much smaller $Q_0^2$, even as low as $Q_0^2=0.16$ GeV$^2$ \cite{Cloet:2007em}.  The best choice of $Q_0^2$ and even the validity of the procedure itself, is, in our view, still an open question.  

The orbital effects which we calculate refer to effective contributions which also include, inevitably, some contributions from glue. Thus our results may not be inconsistent with recent lattice QCD calculations that suggest large contributions from the glue \cite{Syritsyn:2011vk}.  This cannot be discussed further until our model is used to compute the energy-momentum tensor and study the spin sum rule (\ref{eq:spinsum}).  This topic, alluded to very briefly  in Sec.\ \ref{sec:IIG}, requires future study.

We have not done a systematic fit to the data by minimizing $\chi^2$.  If we did so, the fits would certainly improve, and we would be in a better position to assess the uniqueness of our final parameters.  Until this is done, the parameters given in Table \ref{tab:PDpar} must be considered preliminary.

How realistic is such a large D-state probability?  First, it must be emphasized that, because we did not fit all the wave function parameters including $n_P$ and $n_D$ in a single search, it is very possible that a good solution with a smaller D-state exists (and we think this is quite likely).    So, without a systematic $\chi^2$ fit, we do not know precisely how large a D-state probability is required by the data.  Here we present evidence of a large D-state.  For comparison, note that the D-state probability in  $^3$H varies from about 7-9\%, depending on the NN force model used  (see Table V in Ref.\ \cite{Schiavilla:1998je}, for example) and at least four times smaller than our estimate. (The P state probability in $^3$H is also quire small, less than 0.16\%; if this were also increased by the same factor of four as the D-state, it would compare with our result for the nucleon.)  In the three nucleon bound state the D-state arises primarily from  the large tensor force which is a feature of the one pion exchange (OPE) part of the NN potential.  If the  force between two quarks is described by a combination of one gluon exchange (OGE) and a confining interaction (and/or OPE, as some models suggest), and the confining interaction is a mixture of scalar and vector components, then a tensor force can arise from both  the OGE and the vector part of the confining interaction \cite{Carlson:1977vi}.  It may be that this mechanism can produce a large D-state, but Isgur, Karl, and Koniuk \cite{Isgur:1981yz}, using the harmonic oscillator model of Isgur and Karl \cite{Isgur:1979be} find a very much smaller value.  On the other hand, our findings are in line with the results of Refs.\ \cite{Garcilazo:2001ck,Valcarce:2005em} on the importance of higher orbital angular momentum components in the baryon  spectrum, affecting the nucleon mass by about 100 MeV, and producing inversion of the relative positions of positive and negative parity nucleonic excitations (unfortunately the  breaking down of the contributions from each of the different partial waves included in those calculations was not reported).  All that we can say here is that a large D-state can explain the DIS observations.  A dynamical calculation is needed to determine whether or not a large D-state is possible.

If the correct explanation of the DIS data is closer to solution 1 with the expected P-state and a much smaller D-state, then spin puzzle is easily explained by a nucleon wave function with modest angular momentum components.  Otherwise, the spin puzzle leads directly to another puzzle -- what are the dynamical origins of such a large D-state, and what effect does this have on other nucleon observables?

\acknowledgments

This work was partially support by Jefferson Science Associates, 
LLC under U.S. DOE Contract No. DE-AC05-06OR23177.
G.~R.~and M.~T.~P.~want to thank Vadim Guzey 
for helpfull discussions.
This work was also partially financed by the European Union
(HadronPhysics2 project ``Study of strongly interacting matter'')
and by the Funda\c{c}\~ao para a Ciencia e a
Tecnologia, under Grant No.~PTDC/FIS/113940/2009,
``Hadron structure with relativistic models''.  
G.~R.~was supported by the Portuguese Funda\c{c}\~ao para 
a Ci\^encia e Tecnologia (FCT) under Grant  
No.~SFRH/BPD/26886/2006.

\appendix

\section{Calculations of the structure functions}\label{app:sf}

\subsection{Simplifications}

The full quark current is (\ref{eq:qcurrent-2}), but, because the subtraction term $\slashed{q} q_\mu/q^2$ can be reconstructed from the first term,  it is sufficient to calculate the $\gamma_\mu$ contribution (referred to as the {\it unsubtracted\/} contribution) only.  To prove this, note that the full current can be constructed from the operation ${\cal C}$, where
\bea
{\cal C}_\mu{}^{\mu'}\gamma_{\mu'}\equiv \left[g_\mu{}^{\mu'} -\frac{q_\mu q^{\mu'}}{q^2} \right] \gamma_{\mu'}=\gamma_\mu-\frac{\slashed{q}q_\mu}{q^2}\label{eq:A1}
\eea
Since this operation does not depend on the quark or diquark variables, it may be applied after the cross section has been calculated, allowing the construction of the full (conserved) current from the unsubtracted current $\gamma_\mu$.

For example, if the cross section arising from the unsubtracted current  is written in terms of 
a general operator  ${\cal O}$
\bea
\left<\gamma_\mu{\cal O}\gamma_\nu\right>\to W_{\mu\nu}
\label{eq:2.5}
\eea
then the full result is obtained by applying the operation (\ref{eq:A1}) to both sides of Eq.~(\ref{eq:2.5}).  On the l.h.s. we get 
\bea
{\cal C}_\mu{}^{\mu'}{\cal C}_\nu{}^{\nu'}\left<\gamma_{\mu'}{\cal O}\gamma_{\nu'}\right>=\left<\left[\gamma_\mu-\frac{\slashed{q}q_\mu}{q^2}\right]{\cal O}\left[\gamma_\nu-\frac{\slashed{q}q_\nu}{q^2}\right]\right>\qquad
\eea
which is precisely the result for the full current (\ref{eq:qcurrent-2}).  
On the r.h.s. we get
\bea
{\cal C}_\mu{}^{\mu'}{\cal C}_\nu{}^{\nu'}
W_{\mu'\nu'}\to 
&&W_{\mu\nu}-\frac{q_\mu (q^{\mu'}W_{\mu'\nu})}{q^2} -\frac{q_\nu (q^{\nu'}W_{\mu\nu'})}{q^2}
\nonumber\\
&& +\frac{q_\mu q_\nu (q^{\mu'}q^{\nu'}W_{\mu'\nu'})}{q^4}\, ,
\eea
which is conserved.  Furthermore, if any terms proportional to either $q_\mu$ or $q_\nu$ arise from the   unsubtracted calculation, they my be ignored because
\bea
{\cal C}_\mu{}^{\mu'}q_{\mu'}=0,  \label{eq:A5}
\eea
 i.e. they vanish once the full result is constructed.

\subsection{Contributions from squares of S, P, and D-states}

\subsubsection{Isoscalar diquark contribution} \label{sec:I=0S&D}

The only contribution from the D-waves to the isoscalar diquark term comes from the square of the $\psi^{D,0}$.  Together with the S and P-state terms this gives
\bea
W_{\mu\nu}^{I=0}=\frac{ (e_q^2)^0}{8M}\int\!\!\!\! \int_{p'k}&&\Big\{n_S^2\,\psi^2_S(P,k)
\,I_{\nu\mu}^{SS} 
\nonumber\\
&&+n_P^2\,\psi_P^2(P,k)I_{\nu\mu}^{PP}
\nonumber\\
&&+ n_D^2 \, \psi_D^2(P,k)  \, I^{DD}_{\nu\mu}\Big\}\qquad\label{eq:squares}
\eea
where $m_q+\slashed{p}' \simeq \slashed{p}'$, a factor of 1/2 from the wave functions has been moved to the factor multiplying the double integral, a factor of $|\widetilde k^2|\to -\widetilde k^2$ coming from the definition of $\Psi^{D,0}$ is included in $I^{DD}$, and the integral is
\bea
\int\!\!\!\! \int_{p'k}
&\equiv& \int\!\!\!\int\frac{d^3p' d^3k}{(2\pi)^2\,4e_qE_s}\delta^4(p'+k-P-q)\cr
&=&\int\frac{d^4k}{(2\pi)^2}\delta_+(m_q^2-p'^2) \delta_+(m_s^2-k^2)\, .\quad \label{DISint}
\eea
The S and P-state traces are
\bea
I_{\nu\mu}^{SS}&&\equiv   {\rm tr}\Big[\gamma_\nu\,\slashed{p}' \gamma_\mu\,{\cal P}_S\Big]
\nonumber\\
I_{\nu\mu}^{PP}&&\equiv   {\rm tr}\Big[\slashed{\widetilde k}\gamma_\nu\,\slashed{p}' \gamma_\mu\,\slashed{\widetilde k}\,{\cal P}_S\Big]
\eea
where we use the shorthand notation
\bea
{\cal P}_S&=&(M+\slashed{P})(1+\gamma^5\slashed{S})\, .
\eea
The operator ${\cal I}^{DD}$ is obtained by summing over the polarization vectors $\varepsilon$ and $\zeta$ using the identity (which holds for all polarization sums defined  in the fixed-axis representation \cite{Gross:2008zza})
\bea
\Delta_{\alpha\beta}=\sum_\Lambda (\varepsilon_\Lambda)_\alpha (\varepsilon^*_\Lambda)_\beta =-\widetilde g_{\alpha\beta}\, . \label{eq:A13}
\eea
Then, using
\bea
\sfrac3{20}\sum_n G_{\alpha\beta'}&&(\widetilde k,\zeta^*_n)
G^{\alpha}{}_\beta(\widetilde k,\zeta_n) 
=-\sfrac{13}{60}\,\widetilde k^2 \,\widetilde g_{\beta'\beta}-\sfrac{7}{20}\,\widetilde k_{\beta'}\widetilde k_\beta\
\nonumber\\[0.05in]
&&=-\sfrac13 \widetilde k^2\widetilde g_{\beta'\beta}-\sfrac7{20} D_{\beta'\beta}(P,k)\, ,
\qquad
\eea
we obtain (including the factor of $-\widetilde k^2$ referred to above)
\bea
I^{DD}_{\nu\mu}&=&\sfrac9{20}\,\widetilde k^2\sum_{\Lambda n}(\varepsilon_\Lambda)^{\alpha'}(\varepsilon_\Lambda^*)^\alpha G_{\alpha'\beta'}(\widetilde k,\zeta^*_n)
G_{\alpha\beta}(\widetilde k,\zeta_n)
\nonumber\\
&&\qquad\times\sfrac13 {\rm tr}\Big[\widetilde \gamma^{\beta'}\gamma^5\, \gamma_\nu\,\slashed{p}' \gamma_\mu \gamma^5\, \widetilde\gamma^\beta\,{\cal P}_S\Big]
\nonumber\\
&=&-\widetilde k^4 I^g_{\nu\mu}-\sfrac7{20}\,\widetilde k^2 I^D_{\nu\mu}
\, .\qquad
\eea
where we introduce two new standard traces
\bea
I_{\nu\mu}^g&&\equiv \sfrac13  {\rm tr}\Big[\widetilde \gamma_\alpha\,\gamma_\nu\,\slashed{p}' \gamma_\mu\,\widetilde \gamma^\alpha\,{\cal P}_S\Big]
\nonumber\\
I_{\nu\mu}^D&&\equiv D_{\alpha\alpha'} {\rm tr}\Big[\widetilde \gamma^\alpha\,\gamma_\nu\,\slashed{p}' \gamma_\mu\,\widetilde \gamma^{\alpha'}\,{\cal P}_S\Big]\, . \label{eq:I1&I2}
\eea
The trace $I^D_{\nu\mu}$ can be expresed in terms of the others.  Using $\tilde k_\alpha\tilde \gamma^\alpha=\slashed{\tilde k}$,
\bea
I^D_{\nu\mu}=I^{PP}_{\nu\mu}-\widetilde k^2 I^g_{\nu\mu}\, . \label{eq:A14}
\eea

Only the terms ${\cal P}_S\to\slashed{P}+M\gamma^5 \slashed{S}$ contribute to all of these traces, and  $I^{PP}$ and $I^g$ can be simplified by moving one of the $\slashed{\tilde k}$ (or $\tilde \gamma^\alpha$) factors through the operator ${\cal P}_S$, giving 
\bea
I^{PP}_{\nu\mu}&=&-\widetilde k^2{\rm tr}\bigg[\gamma_\nu\,\slashed{p}'\,\gamma_\mu\slashed{P}\Big]+ \widetilde k^2 M \,{\rm tr}\bigg[\gamma_\nu\,\slashed{p}'\,\gamma_\mu\gamma^5\slashed{S}\Big]
\cr
&&-2M\,(S\cdot k)\,{\rm tr}\bigg[\slashed{\tilde k}\;\gamma_\nu\,\slashed{p}'\,\gamma_\mu \gamma^5\bigg]
\nonumber\\
&=&-\widetilde k^2\,I^{SS}_{\mu\nu}-8M(S\cdot k)\,i\epsilon_{\mu\nu\alpha\beta}\,\widetilde k^\alpha p'^\beta\, , \label{eq:A9a} \qquad
\\
I^g_{\nu\mu}&=&-{\rm tr}\bigg[\gamma_\nu\,\slashed{p}'\,\gamma_\mu\slashed{P}\Big]+  M \,{\rm tr}\bigg[\gamma_\nu\,\slashed{p}'\,\gamma_\mu\gamma^5\slashed{S}\Big]
\cr
&&-\sfrac23\,M\,{\rm tr}\bigg[\gamma_\nu\,\slashed{p}'\,\gamma_\mu \gamma^5\,\slashed{S}\bigg]
\nonumber\\
&=& -\sfrac13 I^{SS}_{\nu\mu}-\sfrac23\,I^{SS}_{\mu\nu}\, , \label{eq:A16}
\eea
where the reader should be careful to distinguish $I^{SS}_{\nu\mu}$ from $I^{SS}_{\mu\nu}$.   Putting this all together gives
\bea
I^{DD}_{\nu\mu}=-\sfrac7{20}\,\widetilde k^2 I^{PP}_{\nu\mu}+\sfrac{13}{60}\,\widetilde k^4\Big[I^{SS}_{\nu\mu}+2\,I^{SS}_{\mu\nu}\Big]
\eea

These considerations show that all of the traces depend on only two elementary traces, $I^{SS}$ and $I^{PP}$, with
\bea
I^{SS}_{\nu\mu}&=& I(p',P)- 4M\,i\epsilon_{\mu\nu\alpha\beta}\;p'^\alpha S^\beta
\label{eq:A9}
\eea 
where 
\bea
 I(p',P)&=&{\rm tr}\Big[\gamma_\nu\,\slashed{p}'\,\gamma_\mu \,\slashed{P}\Big]
 \cr &=&
4[p'_\mu P_\nu+P_\mu p'_\nu-g_{\mu\nu}(P\cdot p')].\qquad\quad
 \label{eq:A10}
\eea 
Because $q\to\infty$,  we may replace $p'\to q$ in the last terms of (\ref{eq:A9}) and (\ref{eq:A10}), but not in the terms proportional to $p'_\mu$ or $p'_\nu$ because, through (\ref{eq:A5}), the $q$ contributions to these terms will vanish.  To reduce these we use the identity
\bea
\frac{1}{2\pi}
\int d\Omega_{k_\perp}\, p'_\mu&=&A_1P_\mu +A_2 q_\mu\to A_1 P_\mu +q_\mu \qquad\label{eq:anglep}
\eea
where the nonleading term, $A_1$, given in Eq.~(\ref{eq:A2}), is needed only when the leading $q_\mu$ contribution vanishes.    

To evaluate the last term in (\ref{eq:A9a}), first note that $p'=q+BP-\tilde k$ where $B=[1-(k\cdot P)/M^2]$.  Hence the $p'$ may be replaced by $q+BP$ and we may use identity (\ref{eq:c5}) to reduce it.  In doing the reduction, note that 
\bea
&&-8M (S\cdot k)\,i \epsilon_{\mu\nu\alpha\beta}\;\tilde k^\alpha (q+BP)^\beta
\nonumber\\
&&\quad\to -8MC_1\,i \epsilon_{\mu\nu\alpha\beta} S^\alpha (q+BP)^\beta
\nonumber\\
&&\qquad-8MC_2(S\cdot q)\,i \epsilon_{\mu\nu\alpha\beta} (q-\frac{\nu}{M}P)^\alpha (q+BP)^\beta
\nonumber\\
&&\quad\to8M^2C_1\,{\cal I}_1 -8M^3\nu\,C_2\,{\cal I}_2.\qquad\quad \label{eq:trace2}
\eea
With these simplifications, the two standard traces become
\bea
I^{SS}_{\nu\mu}=&& \,4\Big[2M^2A_1\left(\frac{P_\mu P_\nu}{M^2}\right)-(P\cdot p')g_{\mu\nu}- M^2{\cal I}_1\Big]
\nonumber\\
I^{PP}_{\nu\mu}=&&\,4\Big[- 2\widetilde k^2 M^2 A_1\left(\frac{P_\mu P_\nu}{M^2}\right)+\widetilde k^2(P\cdot p') g_{\mu\nu}
\nonumber\\
&&\qquad-M^2(\widetilde k^2-2\,C_1) {\cal I}_1-2M^3\,\nu\,C_2\,{\cal I}_2\Big]\, . \qquad\label{eq:A9b}
\eea 

Using the relations 
(valid in the DIS limit and derived in Appendix \ref{app:B}) 
\bea
\tilde k^2&=&-{\bf k}^2
\nonumber\\
(P\cdot p')&\to&M\nu
\nonumber\\
A_1&\to& x
\nonumber\\
C_1& = &    -\sfrac12 k_\perp^2=-\sfrac12 {\bf k}^2 (1-z^2) 
\nonumber\\
\tilde k^2-2C_1 & = &  
-k_z^2  = -\sfrac13 {\bf k}^2[1+2\,P_2(z)]
\nonumber\\
\nu^2 C_2&=&c_2\to {\bf k}^2 (\sfrac32 z^2-\sfrac12)={\bf k}^2\,P_2(z) \label{eq:C1&C2}
\eea
where $\widetilde k=\{0, {\bf k}\}$  and $z=\cos\theta$, the three traces 
from Eq.\ (\ref{eq:squares}) reduce to
\bea
I^{SS}_{\nu\mu}=&& \,4M \Big[2Mx\left(\frac{P_\mu P_\nu}{M^2}\right)-\nu g_{\mu\nu}- M{\cal I}_1\Big]
\nonumber\\
I^{PP}_{\nu\mu}=&&\,4M{\bf k}^2\bigg[2 Mx\left(\frac{P_\mu P_\nu}{M^2}\right)-\nu g_{\mu\nu}+\sfrac13 M{\cal I}_1
\nonumber\\
&&\qquad+ P_2(z)\bigg\{\sfrac23M{\cal I}_1-\frac{2M^2}{\nu}{\cal I}_2\bigg\}\bigg]
\nonumber\\
I^{DD}_{\nu\mu}
=&&\,4M{\bf k}^4\bigg\{ \Big[2Mx\left(\frac{P_\mu P_\nu}{M^2}\right)-\nu g_{\mu\nu}+\sfrac{1}{3} M{\cal I}_1\Big]
\nonumber\\
&&\qquad+\sfrac7{20} P_2(z)\Big[\sfrac23M{\cal I}_1-\frac{2M^2}{\nu}{\cal I}_2\Big]\bigg\}
\, . \label{eq:A9bc}
\eea 
These traces are all of the general form
\bea
I^{LL}_{\nu\mu}=8M^2\Big[-z_1 g_{\mu\nu}+z_2\frac{P_\mu P_\nu}{M^2} -z_3{\cal I}_1 + z_4{\cal I}_2\Big],\quad\qquad \label{eq:ziterms}
\eea
Including the factor of $1/(2\pi)$ from (\ref{hadcurr}), the invariants then have the general form (with $I=0$ or 1)
\bea
Z_i^I= (e^2_q)^I \frac{M}{2\pi} \int\!\!\!\! \int_{p'k} z_i\, n_L^2\,\psi_L^2(P,k)  \label{eq:Zi}
\eea
where $Z_i=\{W_1,W_2, \widetilde G_1,G_2\}$ (with $i=\{1,4\}$) and $z_i$ is the coefficient from (\ref{eq:ziterms}).

Combining all of these results  gives
\bea
&&2MxW_1^{I=0}=\nu W_2^{I=0}
\nonumber\\
&&\quad=(e^2_q)^0\frac{2M\nu\,x}{4\pi} \int\!\!\!\! \int_{p'k} 
\Big[n_S^2\,\psi^2_S+n_P^2\,{\bf k}^2\psi_P^2+n_D^2\,{\bf k}^4\psi_D^2\Big]\cr
&&\widetilde G_1^{I=0}=(e^2_q)^0\frac{M}{4\pi}  \int\!\!\!\! \int_{p'k}\Big[n_S^2\,\psi^2_S-\sfrac13\,n_P^2\,{\bf k}^2\psi_P^2-\sfrac13\,n_D^2\,{\bf k}^4\psi_D^2
\nonumber\\
&&\qquad\qquad\qquad-P_2(z)\Big(\sfrac23\,n_P^2\,{\bf k}^2\psi_P^2+\sfrac7{30}\,n_D^2\,{\bf k}^4\psi_D^2\Big)\Big]
\nonumber\\
&&\frac{\nu}{M}G_2^{I=0}=-(e^2_q)^0\frac{M}{2\pi}  \int\!\!\!\! \int_{p'k}P_2(z)
\nonumber\\
&&\qquad\qquad\qquad\times\Big[n_P^2\,{\bf k}^2\psi_P^2+\sfrac7{20}\,n_D^2\,{\bf k}^4\psi_D^2\Big]
\, , \label{eq:W12}
\eea
where $\psi_L=\psi_L(P,k)$, and
\bea
\widetilde G_1\equiv  G_1+\frac{\nu}{M} G_2\, .
\eea
Hence, the isospin zero structure function $G_1$ becomes 
\bea
G_1^{I=0}&=&(e^2_q)^0\frac{M}{4\pi}  \int\!\!\!\! \int_{p'k} \Big[n_S^2\,\psi^2_S-\sfrac13\,n_P^2\,{\bf k}^2\psi_P^2-\sfrac13\,n_D^2\,{\bf k}^4\psi_D^2
\nonumber\\
&&\qquad+P_2(z)\Big(\sfrac43\,n_P^2\,{\bf k}^2\psi_P^2+\sfrac7{15}\,n_D^2\,{\bf k}^4\psi_D^2\Big)\Big] .
\eea

Note that the  $W_1$ and $W_2$ structure functions satisfy the Callen-Gross relation, which will hold for all contributions calculated below.

\subsubsection{Isovector diquark contribution}

Next, look at the contributions from the $I=1$ diquark.  Here there are four contributions:  from the squares of $\Psi^{S,1}$, $\psi^{P,1}$, $\Psi^{D,1}$, and $\Psi^{D,2}$.  All of these involve sums over the diquark polarization $\Lambda$ evaluated using  (\ref{eq:A13}).  Carrying out the spin sum and removing the $\gamma^5$'s (which introduces another minus sign) gives four traces
\bea
W_{\mu\nu}^{I=1}=\frac{(e_q^2)^1}{8M}&&\int\!\!\!\! \int_{p'k}\Big\{n_S^2\,\psi^2_S(P,k)\,J^{SS}_{\nu\mu}
+n_P^2\,\psi_P^2\,J^{PP}_{\nu\mu}
\nonumber\\
&&+n_D^2\,\psi_D^2(P,k)\left(J_{\nu\mu}^{1,1}+J_{\nu\mu}^{2,2}\right)\Big\}\qquad
\eea
where, using
$\widetilde \gamma^\alpha \slashed{P}+
\slashed{P}\,\widetilde \gamma^\alpha =0$ and
\bea
D^{\beta\alpha}D_\beta{}^{\alpha'}= \widetilde k^2 \Big[\sfrac19\, \widetilde k^2 \tilde g^{\alpha\alpha'}+\sfrac13 \widetilde k^\alpha \widetilde k^{\alpha'}\Big],
\eea
these traces are
\bea
J^{SS}_{\nu\mu}&=&-\sfrac13{\rm tr}\Big[\widetilde\gamma_\alpha\,\gamma_\nu\,\slashed{p}'\,\gamma_\mu\,\widetilde\gamma^{\alpha}{\cal P}_S\Big] =-I^g_{\nu\mu}
\nonumber\\
&=&\sfrac13\,I^{SS}_{\nu\mu}+\sfrac23\,I^{SS}_{\mu\nu}
\nonumber\\
J^{PP}_{\nu\mu}&=&-\sfrac13{\rm tr}\Big[\widetilde\gamma_\alpha\,\slashed{\widetilde k}\,\gamma_\nu\,\slashed{p}'\,\gamma_\mu\,\slashed{\widetilde k}\,\widetilde\gamma^{\alpha}{\cal P}_S\Big] 
\nonumber\\
&=&\sfrac13 I^{PP}_{\nu\mu}+\sfrac23 I^{PP}_{\mu\nu}
\nonumber\\
J^{1,1}_{\nu\mu}&=&-\sfrac15\, \widetilde k^4\,I^{g}_{\nu\mu}
\nonumber\\
J^{2,2}_{\nu\mu}&=&-\sfrac65\,D^{\beta\alpha}D_\beta{}^{\alpha'} {\rm tr}\Big[\widetilde\gamma_\alpha\,\gamma_\nu\,\slashed{p}'\,\gamma_\mu\,\widetilde\gamma_{\alpha'}{\cal P}_S\Big]  
\nonumber\\
&=&-\sfrac25 \,\widetilde k^2\,I^{PP}_{\nu\mu}-\sfrac2{5}\,\widetilde k^4\,I^{g}_{\nu\mu}\, .\qquad
\label{eq:A18}
\eea
In the previous section, all of these traces were    expressed in terms of the two standard traces (\ref{eq:A9b}).  Adding the two D-state contributions together, and simplifying, gives
\bea
J^{DD}_{\nu\mu}&=&-\sfrac25 \,\widetilde k^2\,I^{PP}_{\nu\mu}-\sfrac35\,\widetilde k^4\, I^g_{\nu\mu}
\nonumber\\
&=&-\sfrac25 \,\widetilde k^2\,I^{PP}_{\nu\mu}+\sfrac15\,\widetilde k^4\Big[I^{SS}_{\nu\mu}+2\,I^{SS}_{\mu\nu}\Big]
\nonumber\\
&=&4M{\bf k}^4\bigg\{2Mx\Big(\frac{P_\mu P_\nu}{M^2}\Big)-\nu g_{\mu\nu}+\sfrac13 M{\cal I}_1
\nonumber\\
&&\qquad+\sfrac25\,P_2(z)\Big\{\sfrac23M{\cal I}_1-\frac{2M^2}{\nu}{\cal I}_2\Big\}\bigg]
\, .
\eea
Inspection of the results for $J^{SS}$ and $J^{PP}$ shows that contributions to the {\it unpolarized\/} structure functions coming from isovector diquarks are equal to the isoscalar ones, while for the {\it polarized\/} structure functions, $J^{SS}_{\nu\mu}\to -\frac13 I^{SS}_{\nu\mu}$ and $J^{PP}_{\nu\mu}\to - \frac13 I^{PP}_{\nu\mu}$.   Using these arguments, we obtain the following results
\bea
&&2MxW_1^{I=1}=\nu W_2^{I=1}
\nonumber\\
&&\quad=(e^2_q)^1 \frac{M\nu\,x}{2\pi} \int\!\!\!\! \int_{p'k} 
\Big[n_S^2\,\psi^2_S+n_P^2\,{\bf k}^2\psi_P^2+n_D^2\,{\bf k}^4\psi_D^2\Big]\cr
&&\widetilde G_1^{I=1}=-(e^2_q)^1\frac{M}{12\pi}  \int\!\!\!\! \int_{p'k}\Big[n_S^2\,\psi^2_S-\sfrac13\,n_P^2\,{\bf k}^2\psi_P^2+n_D^2\,{\bf k}^4\psi_D^2
\nonumber\\
&&\qquad\qquad\qquad-P_2(z)\Big(\sfrac23\,n_P^2\,{\bf k}^2\psi_P^2-\sfrac45\,n_D^2\,{\bf k}^4\psi_D^2\Big)\Big]
\nonumber\\
&&\frac{\nu}{M}G_2^{I=1}=(e^2_q)^1\frac{2M}{4\pi}  \int\!\!\!\! \int_{p'k}P_2(z)
\nonumber\\
&&\qquad\qquad\qquad\times\Big[\sfrac13n_P^2\,{\bf k}^2\psi_P^2-\sfrac25\,n_D^2\,{\bf k}^4\psi_D^2\Big]
\, , \label{eq:W12a}
\eea
Extracting $G_1$ gives
\bea
G_1^{I=1}=&&-(e^2_q)^1\frac{M}{12\pi}  \int\!\!\!\! \int_{p'k}\Big[n_S^2\,\psi^2_S-\sfrac13\,n_P^2\,{\bf k}^2\psi_P^2+ n_D^2\,{\bf k}^4\psi_D^2
\cr
&&+8 P_2(z)\Big[\sfrac16\,n_P^2\,{\bf k}^2\psi_P^2-\sfrac15 n_D^2\,{\bf k}^4\psi_D^2\Big]\bigg\}.\qquad\quad
\eea

Now add the isospin 0 and isospin 1 parts, separating them into the $u$ and $d$ quark contributions (for the proton) following the discussion which lead to Eq.~(\ref{eq:2.24}).   Collecting terms gives
\bea
&&2MxW^p_1=\nu W_2^p
=x\, \frac{M\nu}{\pi} \int\!\!\!\! \int_{p'k} 
\Big[2\,e_u^2\,{\cal N}_u+e_d^2\,{\cal N}_d\Big] \cr
&&G^p_1=\frac{M}{6\pi}  \int\!\!\!\! \int_{p'k} \bigg\{n_S^2\Big[4\,e_u^2\,(\psi_u^S)^2-e_d^2\,(\psi_d^S)^2\Big]
\nonumber\\
&&\qquad-n_P^2\Big[\sfrac43\,e_u^2\,{\bf k}^2(\psi_u^P)^2-\sfrac13\,e_d^2\,{\bf k}^2(\psi_d^P)^2\Big]
\nonumber\\[0.05in]
&&\qquad-n_D^2\Big[2\,e_u^2\,{\bf k}^4 (\psi_u^D)^2+e_d^2\,{\bf k}^4(\psi_d^D)^2\Big]
\nonumber\\[0.05in]
&&\qquad+n_P^2\,P_2(z)\Big[\sfrac{16}{3}\,e_u^2\,{\bf k}^2 (\psi_u^P)^2-\sfrac{4}3\,e_d^2\,{\bf k}^2(\psi_d^P)^2\Big]
\nonumber\\[0.05in]
&&\qquad+n_D^2\,P_2(z)\Big[\sfrac{29}{10}\,e_u^2\,{\bf k}^4 (\psi_u^D)^2+\sfrac85\,e_d^2\,{\bf k}^4(\psi_d^D)^2\Big]\bigg\}
\nonumber\\
&&\frac\nu{M}G^p_2=-\frac{M}{\pi}  \int\!\!\!\! \int_{p'k}P_2(z)
\nonumber\\
&&\qquad\bigg\{n_P^2\Big[\sfrac43\,e_u^2\,{\bf k}^2(\psi_u^P)^2-\sfrac13\,e_d^2\,{\bf k}^2(\psi_d^P)^2\Big]
\nonumber\\
&&\qquad +n_D^2\Big[\sfrac{29}{40}\,e_u^2\,{\bf k}^4(\psi_u^D)^2+\sfrac25\,e_d^2\,{\bf k}^4(\psi_d^D)^2\Big]\bigg\},\quad
\qquad \label{eq:W12axx}
\eea
where
\bea
{\cal N}_q\equiv n_S^2 (\psi_q^S)^2+n_P^2\,{\bf k}^2(\psi_q^P)^2+n_D^2\,{\bf k}^4(\psi_q^D)^2 \qquad
\eea
is the density factor that appears in the unpolarized structure functions.

\subsection{Extraction of the DIS limit} 

Before calculating the interference terms, we extract the DIS limit of the above results.   The expressions for all of the structure functions are covariant, but it is convenient to evaluate them in the laboratory system, where the two external four-vectors  are
\bea
P&=&\left\{M,0,0,0\right\}\cr
q&=&\left\{\frac{Q^2}{2Mx},0,0,\sqrt{Q^2+\frac{Q^4}{4M^2x^2}}\right\}
\eea
where $x$ is the usual Bjorken scaling variable. 
For the integration of functions of $|{\bf k}|$ and $k_z =|{\bf k}| z$, we 
can write in the DIS limit
\bea
\int\!\!\!\! \int_{p'k}
&=& \int\frac{d^4k}{(2\pi)^2}\delta_+(m_q^2-p'^2) \delta_+(m_s^2-k^2)
\nonumber\\
& =& 
\int \frac{d^3 k}{(2\pi)^2 2 E_s} 
\delta\left(\frac{Q^2}{Mx}[(1-x)-E_s+ |{\bf k}| z ] \right).
\nonumber \\
&=&
\frac{Mx}{Q^2}
\int_0^\infty \frac{ {\bf k}^2 d |{\bf k}|}{
4\pi E_s}\int_{-1}^1 
dz\,\delta\left([M (1-x)-E_s + |{\bf k}| z]\right).
\nonumber
\ea
Scaling all momenta by the nucleon mass 
(so that $|{\bf k}|=M\kappa$) the double integral becomes
\bea
\int\!\!\!\! \int_{p'k}
&=&
\frac{M^2x}{ Q^2}
\int_0^\infty\frac{\kappa d\kappa}{4\pi E_\kappa} 
\int_{-1}^1 \kappa dz\,\delta
\left( (1-x)-E_\kappa +\kappa z\right)
\nonumber\\
&=&
\frac{M^2x}{ Q^2}
\int\frac{\kappa d\kappa}{4\pi E_\kappa}\int_{-1}^1 dz \,\delta\big(z_0-z\big)
\nonumber\\
&=&\frac{m_s}{\nu}\int_{\zeta}^\infty\frac{d\chi}{16\pi},\label{DISint-a}
\eea  
where $\chi$ was defined in Eq.~(\ref{eq:chi}),  $E_\kappa=\sqrt{r^2+\kappa^2}$, $z=z_0$ with $z_0$ defined in (\ref{eq:z0}), and the lower limit of the $\chi$ integral (which occurs at  $|z_0|=1$) was given in Eq.~(\ref{eq:3.13}).  
To obtain the final expression we use
\bea
\int_{\chi} =
\frac{Mm_s}{16\pi^2 }\int_\zeta^\infty d\chi =
\frac{ M \nu}{\pi}
\int\!\!\!\! \int_{p'k}.
\eea

When inserted into (\ref{eq:Zi}), the structure functions have the form
\bea
Z_i^I=(e^2_q)^I \frac{Mm_s}{2\nu(4\pi)^2} \int_\zeta^\infty d\chi \,z_i\, n_L^2\,\psi_L^2(P,k)  \label{eq:Zi2}
\eea
where the $z_i$ are the factors that appear in Eq.\ (\ref{eq:ziterms}).

 For the proton this is
\bea
\nu W^p_2 = 2MxW^p_1&=&x\,\Big[ 2\,e_u^2\,f'_u(x)+e_d^2\,f'_d(x)\Big]
\qquad\quad  \label{eq:B23}
\eea
 where [after the SP interference terms have been calculated, this $f'_q$ is    replaced by the $f_q$ given in Eq.~(\ref{eq:fq})]
 \bea
 f'_q&=&n_S^2f_q^S+n_P^2\,f_q^P+n_D^2\,f_q^D
\eea
 with the individual quark distribution functions $f_q^L(x)$ given by 
\bea
f_q^L(x) =\frac{Mm_s}{(4\pi )^2} \int_{\zeta}^\infty d\chi\,k^{2\ell}[\psi^L_q(\chi)]^2, \label{eq:A37a}
 \eea
where now $k=|{\bf k}|$, and $L=\{S, P, D\}$ with $k^{2\ell}=1, k^2,$ or $k^4$ for S, P, or D-states, respectively. 
Up to the normalization factor, treated differently in this paper, the structure functions $f_q^S(x)$ were already obtained in Ref.~\cite{Gross:2006fg}.   The interpretation of Eq.\ (\ref{eq:A37a}) was discussed in Sec.\ \ref{sec:IIIA} above. 

The other proton structure functions become
\bea
\nu G^p_1&&\,=g_1^p(x)=n_S^2\sfrac16[4\,e_u^2\,f_u^S(x)-e_d^2\,f_d^S(x)]
\nonumber\\
&&\qquad\quad-n_P^2\sfrac1{18}[4\,e_u^2\,f_u^P(x)-e_d^2\,f_d^P]
\nonumber\\
&&\qquad\quad-n_D^2\sfrac1{6}[2\,e_u^2\,f_u^D(x)+e_d^2\,f_d^D(x)]
\nonumber\\[0.05in]
&&\qquad\quad+n_P^2\,\sfrac29[4\,e_u^2\,g^P_u(x)-e_d^2\,g^P_d(x)]
\nonumber\\[0.05in]
&&\qquad\quad+n_D^2\sfrac1{60}[29\,e_u^2\,g^D_u(x)+16\,e_d^2\,g^D_d(x)]
\nonumber\\
\frac{\nu^2}{M}G^p_2\,&&=g_2^p(x)=
-n_P^2\,\sfrac13[4\,e_u^2\,g^P_u(x)-e_d^2\,g^P_d(x)]
\nonumber\\
&&\qquad\quad-n_D^2\sfrac1{40}[29\,e_u^2\,g^D_u(x)+16\,e_d^2\,g^D_d(x)]\qquad\quad \label{eq:335}
\eea
where the new structure functions, both of which depend on $P_2(z_0)$, are
\bea
g^P_q(x)&=&\frac{Mm_s}{(4\pi )^2} \int_{\zeta}^\infty d\chi\,P_2(z_0)\,k^{2}[\psi^P_q(\chi)]^2
\nonumber\\
g^D_q(x)&=&\frac{Mm_s}{(4\pi )^2} \int_{\zeta}^\infty d\chi\,P_2(z_0)\,k^{4}[\psi^D_q(\chi)]^2,
\quad \label{eq:A37b}
\eea
where $z_0$ was defined in Eq.\ (\ref{eq:z0}).

\subsection{DIS limit in light cone variables}

The DIS limit can also be evaluated in light-cone coordinates. In our notation, an arbitrary four-vector $v$ is written
\bea
v=\{v_+,v_-,{\bf v}_\perp\}\qquad
v_\pm = v_0\pm v_3
\eea
so that the scalar product is
\bea
v\cdot u = v_\mu u^\mu=\sfrac12(v_+u_-+v_-u_+)-{\bf v}_\perp\cdot{\bf u}_\perp\, .
\eea
In the DIS limit, the four-vectors $P$ and $q$, in light-cone notation in the laboratory frame are
\bea
P&=&\{M,M,{\bf 0}\}\cr
q&=&\left\{\frac{Q^2}{Mx}, -Mx, {\bf 0}\right\}\, .
\eea

Using this notation the double integral can be reduced to an integral over $k_\perp^2$.  Defining $k_-=M(1-y)$, the integral becomes 
\bea
\int\!\!\!\! \int_{p'k}
&=& \int\frac{d^4k}{(2\pi)^2}\delta_+(m_q^2-p'^2) \delta_+(m_s^2-k^2)
\nonumber\\
&\to&\int\frac{dk_\perp^2}{4\pi}\frac12\int dk_-\!\int dk_+\,\delta\big(m_s^2+k_\perp^2-k_+k_-\big)
\nonumber\\
&&\qquad\times\delta\left(\frac{Q^2}{Mx}[M+q_--k_-]\right)
\nonumber\\
&=&\int\frac{dk_\perp^2}{8\pi}\int \frac{dy}{(1-y)}\,\frac{x}{Q^2}\,\delta\big(y-x\big)
\nonumber\\
&=&\frac{1}{M\nu}\int\frac{dk_\perp^2}{16\pi(1-x)},\label{DISint-ab}
\eea
where the $\delta(y-x)$ fixes $k_-=M(1-x)$.  

In light-cone variables the rotational symmetry is broken so the variable $\chi$, which previously depended only on $\kappa$, now depends on both $x$ and $k_\perp^2\equiv Mm_s\eta$ 
\bea
\chi\to\chi_{_{\rm LC}}&&=\chi(\eta,x)=\frac{k_++k_-}{m_s}-2
\nonumber\\
&& =\frac {r+\eta}{1-x} +\frac{1-x}{r}-2
=\zeta+\frac{\eta}{1-x}\qquad\quad \label{eq:chiLC}
\eea
However, the integral (\ref{DISint-ab}) can be immediately transformed into (\ref{DISint-a}), showing that the representations are equivalent.    We used the light cone form (\ref{DISint-ab}) in Ref.~\cite{Gross:2006fg}, but prefer (\ref{DISint-a}) for this paper because of the interpretation of its meaning, as discussed in Sec.\ \ref{sec:IIIA} above. 

In order to finish the comparison, note that, in light cone variables, the weight functions that appear in (\ref{kpinta}) are
\bea
&&k\,z_0= k_z=\sfrac12(k_+-k_-)=\sfrac12\,m_s\Big[\chi_{_{\rm LC}}+2\left(1-\frac{y}{r}\right)\Big]
\nonumber\\
&&k^2(1-z_0^2)=k_\perp^2 =Mm_sy(\chi-\zeta)
\nonumber\\[0.05in]
&&k^2\,P_2(z_0)=k_z^2-\sfrac12\,k_\perp^2\, ,
\eea
where, as elsewhere, $y=1-x$.

\subsection{Interference Terms}

\subsubsection{SP interference}

The contribution to the SP interference term from isoscalar diquarks is 
\bea
W_{\mu\nu}^{I=0}=\frac{ (e_q^2)^0}{8M}\int\!\!\!\! \int_{p'k}n_Pn_S\,\psi_{S}(P,k)\psi_P(P,k)I^{SP}_{\nu\mu}\qquad
 \eea
where the new trace is
\bea
I^{SP}_{\nu\mu}= {\rm tr}\Big[\slashed{\tilde k}\gamma_\nu\slashed{p}'\gamma_\mu {\cal P}_S+\gamma_\nu\slashed{p}'\gamma_\mu \slashed{\tilde k}{\cal P}_S\Big]
\eea     
and we have been careful to separately write the two contributions from the overlap of the final P-state (the first term) and the initial P-state (the second term).  As it turns out, these terms are not identical.  To reduce the calculation, note that the trace now picks up the terms with an even number of $\gamma$ matrices in ${\cal P}_S$
\bea
{\cal P}_S\to M+\slashed{P}\gamma^5\slashed{S}. \label{eq:A37}
\eea
Moving the $\slashed{\tilde k}$ through ${\cal P}_S$ reduces the trace to
\bea
I^{SP}_{\nu\mu}
&=&2M{\rm tr}\Big[\slashed{\tilde k}\gamma_\nu\slashed{p}'\gamma_\mu\Big] +2(\tilde k\cdot S){\rm tr}\Big[\gamma_\nu\slashed{p}'\gamma_\mu\slashed{P}\gamma^5\Big] \qquad
\nonumber\\
&=&8M[\tilde k_\nu p'_\mu+p'_\nu \tilde k_\mu-(\tilde k\cdot p')g_{\mu\nu}\Big]
\nonumber\\
&&+8i\,(\tilde k\cdot S)\,\epsilon_{\mu\nu\alpha\beta}\,p'^\alpha P^\beta .
\eea
In order to expand this into the four independent DIS structure functions, we recall that $p'=q+BP-\tilde k$ (with $MB=M-E_s$), and average over the directions of ${\bf k}_\perp$ using the identities (\ref{eq:C2}) and (\ref{eq:c5}).  We get
\bea
I^{SP}_{\nu\mu}&=&8M\Big[B_1\big(\tilde q_\nu (q+BP)_\mu+(q+BP)_\nu \tilde q_\mu\Big)
\nonumber\\
&&\qquad \quad-2C_1\tilde g_{\mu\nu}-2C_2\tilde q_\mu \tilde q_\mu-(\tilde k\cdot p')g_{\mu\nu}\Big]
\nonumber\\
&&+8M^3\,(B_1-C_2)\,{\cal I}_2
+8i\,C_1\,\epsilon_{\mu\nu\alpha\beta}\,P^\alpha S^\beta\qquad
\eea
Dropping all terms proportional to to $q_\mu$ and $q_\nu$, the unpolarized terms reduce to 
\bea
I^{SP}_{\nu\mu}\Big|_u=&&-8M\,g_{\mu\nu}[2C_1+ (\tilde k\cdot p')]
\nonumber\\
&&-8M\,\frac{P_\mu P_\nu}{M^2}\Big[2M\nu B_1B-2(C_1-\nu^2 C_2)\Big]\qquad
\eea
and in the DIS limit the coefficients  can be simplified:
\bea
&&2C_1+(\tilde k\cdot p')\to-k_\perp^2 -k_z q_z +k^2\to -k_z\nu
\nonumber\\
&&2M\nu BB_1-2(C_1-C_2\nu^2)\to 2(M-E_s)k_z +2k_z^2
\nonumber\\
&&\qquad\to 2k_z(k_z+M-E_s)=2k_zMx\, .
\eea
where the light-cone relation $k_-=E_s-k_z=M(1-x)$ has been used.  Substituting, gives
\bea
I^{SP}_{\nu\mu}\Big|_u=&&8M\Big[k_z\,\nu\,g_{\mu\nu}
-2\,k_z\,M\,x\left(\frac{P_\mu P_\nu}{M^2}\right)\Big]\qquad
\eea

The polarized terms  contain an invariant not in the canonical form.  It can be re-expressed using the identity (\ref{eq:A4}), giving
\bea
I^{SP}_{\nu\mu}\Big|_p&=&{\cal I}_1(8C_1E_1)
\nonumber\\
&&+{\cal I}_2[8M^3(B_1-C_2)+8C_1E_2].\qquad
\eea
In the DIS limit these coefficients become 
\bea
&&8C_1E_1\to 8M\,\frac{k_\perp^2}{4x} 
\nonumber\\
&&8M^3(B_1-C_2)+8C_1E_2
\nonumber\\
&&\qquad\to 8M^3\left(\frac{k_z}{\nu}-\frac1{\nu^2}\Big[k_z^2-\sfrac12 k_\perp^2\Big] \right)
-4M^3\frac{k_\perp^2}{Q^2}\qquad
\nonumber\\
&&\qquad\to \frac{8M^2}{\nu}\Big[Mk_z-\frac{k_\perp^2}{4x}\Big]
\eea
giving
\bea
I^{SP}_{\nu\mu}\Big|_p&=&8M^2\Big[\frac{k_\perp^2}{4Mx} {\cal I}_1 +\frac{M}{\nu}\left(k_z-\frac{k_\perp^2}{4Mx}\right){\cal I}_2\Big].\qquad \label{eq:SPpol}
\eea
 
Using (\ref{eq:ziterms}) and (\ref{eq:Zi2}), the isospin zero contribution to the SP interference term is then
\bea
\nu W_2^{I=0}&&=2Mx W_1^{I=0}
=-n_Pn_S\,(e_q^2)^0x\,h_q^0(x) 
\nonumber\\
 \nu\,\tilde G_1^{I=0}&&=-n_Pn_S\,\sfrac12\,(e_q^2)^0 \,h_q^0(x)
 \nonumber\\
\frac{\nu^2}{M}\,G_2^{I=0}&&=-n_Pn_S\,\sfrac12\,(e_q^2)^0\Big[h_q^1(x)-h_q^0(x)\Big],\qquad
\eea
where we introduce two new structure functions involving the overlap of the S and P-states
\bea
h_q^0(x)&&=\frac{Mm_s}{(4\pi)^2}\int_\zeta^\infty d\chi\, k \,z_0\, \psi_q^{S}(\chi)\psi_q^{P}(\chi)
\nonumber\\
h_q^1(x)&&=\frac{Mm_s}{(4\pi)^2}\int_\zeta^\infty d\chi\,  \frac{k^2(1-z_0^2)}{4Mx}\psi_q^{S}(\chi)\psi_q^{P}(\chi).
\qquad\quad \label{eq:hfunctions}
\eea

It is now an easy matter to compute the isovector contribution to the SP interference term 
%
\bea
W_{\mu\nu}^{I=1}&&=\frac{ (e_q^2)^1}{8M}\int\!\!\!\! \int_{p'k} n_S n_P\,\psi_S(P,k)\psi_P(P,k)\,J^{SP}_{\nu\mu}\qquad\quad
 \eea
 where, removing the $\gamma^5$'s and summing over the diquark polarization, gives 
 \bea
J^{SP}_{\nu\mu}&=&\sfrac13{\rm tr}\Big[(\widetilde\gamma_\alpha\,\slashed{\tilde k}\gamma_\nu\slashed{p}'\gamma_\mu \,\widetilde\gamma^\alpha
+\widetilde\gamma_\alpha\,\gamma_\nu\slashed{p}'\gamma_\mu \slashed{\tilde k}\,\widetilde\gamma^\alpha){\cal P}_S\Big].\qquad
\eea   
Using the identity
\bea
\widetilde\gamma_\alpha (M+\slashed{P}\gamma^5\slashed{S})\widetilde\gamma^\alpha=3M-\slashed{P}\gamma^5\slashed{S}
\eea
we see immediately that the unpolarized part of $J^{SP}$ equals the unpolarized part of $I^{SP}$, while the polarized part of $J^{SP}$ is $-\frac13$ of the polarized part of $I^{SP}$.   This can be be summarized by the relation
\bea
J^{SP}_{\nu\mu}&=&\sfrac13 I^{SP}_{\nu\mu}+\sfrac23 I^{SP}_{\mu\nu}
\eea
previously encountered for $J^{SS}$ and $J^{PP}$ in Eq.\ (\ref{eq:A18}).

Adding the two isospin  contributions, and separating the $u$ and $d$ quarks, gives the following result for the proton structure functions 
\bea
\nu W_2^p(x)&&=2MxW_1^p=-2n_Pn_S\,x[2\,e_u^2\,h_u^0(x)+e_d^2\,h_d^0(x)]
\nonumber\\
g_1^p(x)&&=-n_Pn_S[\sfrac43\,e_u^2 h_u^0(x)-\sfrac13\,e_d^2\,h_d^0(x)]
\nonumber\\
g_2^p(x)&&=-n_Pn_S\Big(\sfrac43\,e_u^2[h_u^1(x)-h_u^0(x)] 
\nonumber\\
&&\qquad\qquad-\sfrac13\,e_d^2[h_d^1(x)-h_d^0(x)]\Big)\, . \label{eq:B42}
\eea
These are combined with the S and D-state contributions as reported in Sec.\ \ref{sec:IID}.

\subsubsection{SD interference}

Only the $\Psi^{D,2}$ D-state component can interfere with the S-state components, and this involves an isovector diquark, giving
\bea
W_{\mu\nu}^{I=1}=\frac{ (e_q^2)^1}{8M}\int\!\!\!\! \int_{p'k}a_{SD}\,\psi_{S}(P,k)\psi_D(P,k)J^{SD}_{\nu\mu}\qquad
 \eea
where $a_{SD}$, defined in Eq.\ (\ref{eq:ad}), includes some factors from the D-state wave function.   Summing over the polarizations of the diquark and removing the $\gamma^5$'s, gives the new trace
\bea
&&J^{SD}_{\nu\mu}=-\sfrac23\,D ^{\alpha\alpha'}{\rm tr}\bigg[\widetilde\gamma_\alpha\,\gamma_\nu\,\slashed{p}'\,\gamma_\mu\, \widetilde\gamma_{\alpha'}
{\cal P}_S\bigg] =-\sfrac23\,I^D_{\nu\mu}\qquad
\eea
where the two identical terms (with the D-state in either the initial or final state) have been combined, and the trace $I^D$ was encountered before, Eq.\ (\ref{eq:I1&I2}).  Using (\ref{eq:A14}), (\ref{eq:A9a}), and  (\ref{eq:A16}),  
\bea
I^D_{\nu\mu}&=&\sfrac13\,\widetilde k^2\Big[ I^{SS}_{\nu\mu}-I^{SS}_{\mu\nu}\Big]-8M(S\cdot k)\,i\epsilon_{\mu\nu\alpha\beta}\,\widetilde k^\alpha p'^\beta\qquad
\nonumber\\
&=&8M^2\Big[\left(C_1-\sfrac13\widetilde k^2\right) {\cal I}_1-M\nu C_2{\cal I}_2\Big]
\nonumber\\
&\to& 8M^2\,{\bf k}^2\,P_2(z)\Big[\sfrac13\,{\cal I}_1-\frac{M}{\nu}{\cal I}_2\Big]
\eea
This contributes
\bea
 \nu\,\tilde G_1^{I=0}&&=\sfrac19\,a_{D}\,(e_q^2)^1 \,d_q(x)
 \nonumber\\
\frac{\nu^2}{M}\,G_2^{I=1}&&=\sfrac13\,a_{D}\,(e_q^2)^1d_q(x),\qquad
\eea
where the new structure function is
\bea
 d_q(x)&&\equiv \frac{Mm_s}{(4\pi)^2}\int_\zeta^\infty d\chi\, P_2(z_0)k^2\,\psi_q^S(\chi)\psi_q^D(\chi). \qquad
\eea
The proton structure functions are
\bea
g_1^p(x)&=&-\sfrac29\,a_{D}\Big[e_u^2d_u(x)+2\,d_d(x)\Big]
\nonumber\\
g_2^p(x)&=&\sfrac13\,a_{D}\Big[e_u^2d_u(x)+2\,d_d(x)\Big]\, . \label{eq:sdinter}
\eea

\subsubsection{PD interference}

As for the SD interference, only the $\Psi^{D,2}$ term will interfere with the P-state, and the trace is
\bea
W_{\mu\nu}^{I=1}=\frac{(e_q^2)^1}{8M} \int\!\!\!\! \int_{p'k} a_{PD}\,\psi_D(p,k)\psi_P(P,k)\, J^{PD}_{\mu\nu}\qquad
\eea  
where $a_{PD}$, defined in Eq.\ (\ref{eq:ad}), includes some factors from the D-state wave function.   Summing over the spin of the diquark and removing the $\gamma^5$'s  gives
\bea
J^{PD}_{\mu\nu}&=&\sfrac13D ^{\alpha\alpha'}{\rm tr}\bigg[\Big(\widetilde\gamma_\alpha\,\gamma_\nu\,\slashed{p}'\,\gamma_\mu\,\slashed{\tilde k}\, \widetilde\gamma_{\alpha'}
+\widetilde\gamma_\alpha\,\slashed{\tilde k}\,\gamma_\nu\,\slashed{p}'\,\gamma_\mu\, \widetilde\gamma_{\alpha'}\Big)
{\cal P}_S\bigg] 
\nonumber\\
&=&\sfrac13\tilde k^2\,{\rm tr}\bigg[\Big(\slashed{\tilde k}\,\gamma_\nu\,\slashed{p}'\,\gamma_\mu+\gamma_\nu\,\slashed{p}'\,\gamma_\mu\, \slashed{\tilde k} \Big) {\cal P}_S
\nonumber\\
&&-\sfrac13\Big(\widetilde\gamma_\alpha\,\gamma_\nu\,\slashed{p}'\,\gamma_\mu\,\slashed{\tilde k}\, \widetilde\gamma^{\alpha}
+\widetilde\gamma_\alpha\,\slashed{\tilde k}\,\gamma_\nu\,\slashed{p}'\,\gamma_\mu\, \widetilde\gamma^{\alpha}\Big)
{\cal P}_S\bigg] 
\eea
where, in this term, ${\cal P}_S\to M+\slashed{P}\gamma^5\slashed{S}$.  The unpolarized terms cancel, and  using the identities 
\bea
\slashed{\tilde k}\,\slashed{P}\gamma^5\slashed{S}+\slashed{P}\gamma^5\slashed{S}\,\slashed{\tilde k}&=&2(\tilde k\cdot S) \slashed{P}\gamma^5
\nonumber\\
\slashed{\widetilde k}\,\widetilde\gamma_\alpha \slashed{P}\gamma^5\slashed{S}\widetilde\gamma^\alpha+ \widetilde\gamma_\alpha \slashed{P}\gamma^5\slashed{S}\widetilde\gamma^\alpha\,\slashed{\widetilde k}&=&-2(\tilde k\cdot S) \slashed{P}\gamma^5
\eea
this trace reduces to
\bea
J^{PD}_{\mu\nu}&=&\sfrac89\tilde k^2\,(\tilde k\cdot S){\rm tr}\Big[\gamma_\nu\,\slashed{p}'\,\gamma_\mu\,\slashed{P}\gamma^5\Big] 
\to
- \sfrac49\,{\bf k}^2\,I^{SP}_{\nu\mu}\Big|_p \qquad
\eea
where $I^{SP}|_p$ was given above, Eq.\ (\ref{eq:SPpol}).  This term contributes to both $\tilde G_1$ and $G_2$, giving
\bea
 \nu\,\tilde G_1^{I=1}&&=\sfrac{2}{9}\,a_{PD}\,(e_q^2)^1 \,h_q^2(x)
 \nonumber\\
\frac{\nu^2}{M}\,G_2^{I=1}&&=\sfrac29\,a_{PD}\,(e_q^2)^1\Big[h_q^3(x)-h_q^2(x)\Big],\qquad
\eea
where the result has been expressed in terms of the new structure functions
\bea
h_q^2(x)&&=\frac{Mm_s}{(4\pi)^2}\int_\zeta^\infty d\chi\, k^3 \,z_0\, \psi_q^{D}(\chi)\psi_q^{P}(\chi)
\nonumber\\
h_q^3(x)&&=\frac{Mm_s}{(4\pi)^2}\int_\zeta^\infty d\chi\,  \frac{k^4(1-z_0^2)}{4Mx}\psi_q^{D}(\chi)\psi_q^{P}(\chi).
\qquad\quad \label{eq:hfunctions2}
\eea

Evaluating this for the proton gives
\bea
g_1^p(x)&&=\sfrac29\,a_{PD}[e_u^2 h_u^2(x)+2\,e_d^2\,h_d^2(x)]
\nonumber\\
g_2^p(x)&&=\sfrac29\,a_{PD}\Big(e_u^2[h_u^3(x)-h_u^2(x)] 
\nonumber\\
&&\qquad\qquad+2\,e_d^2[h_d^3(x)-h_d^2(x)]\Big)\, . \label{eq:B42a}
\eea
Note that the structure of thess terms is similar to the SP interference terms, except for the charge weightings, which parallel the SD interference terms.

\section{Details in the derivation of the DIS formulae} \label{app:B}

First, evaluate the average of $\tilde k_\alpha$ over the directions of ${\bf k}_\perp$.  Using the fact that the wave functions are independent of the direction of ${\bf k}_\perp$, and that $P\cdot \tilde k=0$, we use the fact that the integral can depend only on 
\bea
\tilde q=q-\frac{(P\cdot q)P}{M^2}=q-\frac{\nu P}{M}\, .
\eea
which gives
\bea
\frac1{2\pi}\int d\Omega_{k_\perp} \tilde k_\alpha&&= B_1\, \tilde q_\alpha\, . \label{eq:C2}
\eea
The coefficient $B_1$ is therefore
\bea
B_1=\frac{\tilde k\cdot\tilde q}{\tilde q^2}=\frac{M(k\cdot q)-\nu(k\cdot P)}{M\tilde q^2}=\frac{k_z}{|{\bf q}|}\to \frac{k_z}{\nu},\qquad
\eea
where the last experssion is the result in the DIS limit.
The average over $p'_\alpha=q+P-k=q+BP-\tilde k$ [where $B=1-(k\cdot P)/M^2$] follows immediately
\bea
\frac{1}{2\pi}\int d\Omega_{k_\perp}p'_\alpha&&=q_\alpha+BP_\alpha-B_1\tilde q_\alpha
\nonumber\\
&&=A_1P_\alpha+A_2q_\alpha \label{eq:C1}
\eea
where $A_1$ and $A_2$ are
\bea
A_1&&=B+B_1\frac{\nu}{M}\to1-\frac{k_-}{M}=x
\nonumber\\
A_2&&=1-B_1\to 1  \label{eq:A2}
\eea

The second integral we encounter is the average
\bea
\frac1{2\pi}\int d\Omega_{k_\perp}\, \tilde k_\alpha \tilde k_\beta&=&C_1\tilde g_{\alpha\beta} +C_2 \tilde q_\alpha \tilde q_\beta
\label{eq:c5}
\eea
where $\tilde g$ was defined in Eq.~(\ref{eq:2.15}).  The simple form (\ref{eq:c5}) follows from the conditions that the contraction of $P$ into either index must give zero.  The coefficients are found from the relations
\bea
\tilde k^2&=&3C_1+C_2 \tilde q^2
\cr
(q\cdot \tilde k)^2&=&C_1 \tilde q^2+C_2\tilde q^4
\eea
giving
\bea
C_1&=&\frac{\tilde q^2\tilde k^2-(q\cdot\tilde k)^2}{2\tilde q^2}
\cr
C_2&=&\frac{3(q\cdot\tilde k)^2-\tilde q^2\tilde k^2}{2\tilde q^4}\, .
\eea
Noting that $\tilde q^2= -{\bf q}^2$ and 
$\tilde k= - {\bf k}^2$ we can write
\bea
C_1&= & -\sfrac12 k_\perp^2=-\sfrac12 {\bf k}^2\sin^2\theta
\nonumber\\
C_2&=& \frac1{{\bf q}^2}\Big[\sfrac32 k_z^2-\sfrac12 {\bf k}^2\Big]=
\frac1{{\bf q}^2}
\Big[ k_z^2-\sfrac12 {k}_\perp^2\Big].
\eea
In the DIS limit
\bea
\tilde q^2&=&-\Big(Q^2+\frac{(P\cdot q)^2}{M^2}\Big)\to-\nu^2,
\eea
we may write
\bea
C_2&\to&\frac1{\nu^2}\Big[ k_z^2-\sfrac12 {k}_\perp^2\Big]
\, .\label{eq:c10}
\eea
To express $C_2$ in terms of $k_\perp^2$, we note that
\bea
k_-&=&M(1-x)
\cr
&=&E_s-k_z=\sqrt{m_s^2+k_\perp^2+k_z^2}-k_z \, .
\eea
Solving for $k_z$ gives
\bea
k_z=\frac{m_s^2+k_\perp^2}{2M(1-x)}-\sfrac12 M(1-x)\, ,
\eea
and hence 
%
\bea
C_2&\to&\frac{1}{\nu^2}\Big[\frac{(m^2_s+k^2_\perp)^2}{4M^2(1-x)^2}-\sfrac12(m_s^2+2k^2_\perp)
+\sfrac14M^2(1-x)^2\Big]
\nonumber\\
&\equiv&\frac{1}{\nu^2}\,c_2. \label{eq:c13}
\eea
Even though $C_2$ is very small, it cannot be neglected because it multiplies a large term.

Note that
\bea
\tilde k^2-3C_1=C_2\,\tilde q^2\to -c_2 = - k_z^2+ \sfrac{1}{2} k_\perp^2 .
\eea

\section{Identity for the reduction of the hadronic tensor}

Here we prove an identity needed for the calculation of the P-state contributions to the hadronic tensor.  This is
\bea
&&i\epsilon_{\mu\nu\alpha\beta}\,P^\alpha S^\beta-\frac{q_\mu}{q^2}i\epsilon_{\lambda\nu\alpha\beta}\,q^\lambda P^\alpha S^\beta -\frac{q_\nu}{q^2}i\epsilon_{\mu\lambda\alpha\beta}\,q^\lambda P^\alpha S^\beta
\nonumber\\
&&\qquad\quad=E_1\,{\cal I}_1 +E_2\,{\cal I}_2 \, .
\eea

To prove this identity note that both sides conserve current and are antisymmetric; therefore the coefficients $E_i$ can be determined by requiring the the projections $P^\mu\,T_{\mu\nu}$ and $S^\mu\,T_{\mu\nu}$ be equal.   This gives
\bea
E_1&&=-\frac{M}{2x}
\nonumber\\
E_2&&=\frac{M^3}{Q^2}=\frac{M^2}{2\nu\,x}\, .
\eea

In our calculation we were able to ignore terms proportional to $q_\mu$ or $q_\nu$.  If these terms are dropped, the identity becomes
\bea
i\epsilon_{\mu\nu\alpha\beta}\,P^\alpha S^\beta&&\to E_1{\cal I}_1 + E_2{\cal I}_2. 
\label{eq:A4}
\eea

\end{document}